\documentclass[11pt]{article}
 
\usepackage{amsfonts,amssymb,amsmath,amsthm,amscd,latexsym}

\def\be{\begin{equation}}
\def\ee{\end{equation}}
\def\bea{\begin{eqnarray}}
\def\eea{\end{eqnarray}}
\newcommand{\ba}{\begin{array}}
\newcommand{\ea}{\end{array}}

\newcommand{\sect}[1]{\setcounter{equation}{0}\section{#1}}
\newcommand{\subsect}[1]{\subsection{#1}}

 \newcommand\dd{{\rm d}}

 \newcommand{\RR}{\mathbb{R}}

  \newcommand{\sss}{S}
\newcommand{\mm}{M}

\newcommand{\xxi}{\tilde \xi}

\def\k{\omega}
\def\>#1{{\bf #1}}                 

 \def\m{{\eta}}  
\def\C{{\Upsilon}}

\newcommand{\Sz}{{\rm\ \!S}_z}            
\newcommand{\Cz}{{\rm\ \!C}_z}

\begin{document}

\  \bigskip
   
\ \bigskip\bigskip

\begin{center}
\baselineskip 24 pt {\Large \bf  
From Lorentzian to Galilean  (2+1) gravity: Drinfel'd doubles, quantisation and noncommutative spacetimes}
\end{center}

\medskip

\begin{center}

{\sc \'Angel  Ballesteros, Francisco J. Herranz  and
Pedro Naranjo}

{Departamento de F\1sica, Universidad de Burgos, 
E-09001 Burgos, Spain}

e-mail: {angelb@ubu.es, fjherranz@ubu.es,  pnaranjo@ubu.es}

\end{center}

\begin{abstract}
It is shown that the canonical classical $r$-matrix arising from the Drinfel'd double structure underlying the two-fold centrally extended (2+1) Galilean and Newton-Hooke Lie algebras (with either zero or non-zero cosmological constant $\Lambda$, respectively) originates as a well-defined non-relativistic contraction of a specific class of canonical $r$-matrices associated with the Drinfel'd double structure of the (2+1) (anti)-de Sitter Lie algebra. The full quantum group structure associated with such (2+1) Galilean and Newton-Hooke Drinfel'd doubles is presented, and the corresponding noncommutative spacetimes are shown to contain a commuting `absolute time' coordinate ${\hat x}_0$ together with two noncommutative space coordinates $({\hat x}_1,{\hat x}_2)$, whose commutator is a function of the cosmological constant $\Lambda$ and of the (central) `quantum time' coordinate ${\hat x}_0$. Thus, the Chern-Simons approach to Galilean (2+1) gravity can be consistently understood as the appropriate non-relativistic limit of the Lorentzian theory, and their associated quantum group symmetries (which do not fall into the family of so-called kappa-deformations) can also be derived from the (anti)-de Sitter quantum doubles through a well-defined quantum group contraction procedure.
\end{abstract}

\medskip 

\noindent
PACS:   02.20.Uw \quad  04.60.-m

\noindent
KEYWORDS:  (2+1) gravity, noncommutative  spacetime, non-relativistic limit,\\ Galilean spacetimes, cosmological constant, quantum groups, Poisson--Lie groups.


\sect{Introduction}

It is well-known that (2+1) gravity proves itself an interesting arena on which the conceptual and technical difficulties of (3+1) gravity are rendered more  amenable. In particular, (2+1) gravity can be described as a Chern-Simons (CS) gauge theory, in which  the gauge group  is the isometry group of the corresponding constant curvature model spacetime \cite{AT, Witten1}.

In this context, quantum groups \cite{CP,majid} have arised as highly promising candidates for symmetries of (2+1) quantum gravity, since the quantum deformations of the isometry groups of the associated model spacetimes (see \cite{Woronowicz, CGSTe3, kappaP, BHOS, Kowalski-Glikman:2002we, amel, symmetry} and references therein) are well-founded  mathematical structures that correspond to the quantisations of the Poisson-Lie (PL) groups defining the phase space of the classical theory in a CS setting and, as a consequence, they could provide a sound (noncommutative) geometric interpretation of the combinatorial or Hamiltonian quantisation of the theory \cite{AGSI,AS,BR,BNR,we2}. 

 We recall (see~\cite{we2,Stalk,Bernd1,Bernd2})   that,  within such combinatorial approach to the quantisation of (2+1) gravity as a CS gauge theory,  an essential role is played by the quantisation of the Fock-Rosly (FR) extended symplectic structure, arising from a compatible classical $r$-matrix (in a sense to be made precise below), defined on the Lie algebra of the isometry group of the model spacetime. Such an extended phase space for a CS theory on $\RR\times \Sigma_{gn}$ ($\Sigma_{gn}$ being  a surface of genus $g$ with $n$ punctures representing particles) turns out to be isomorphic to the Cartesian product of $g$ copies of the Heisenberg double Poisson structure associated with the same $r$-matrix and $n$ copies of a symplectic leaf of the dual PL group structure on the gauge group, which is also defined in terms of $r$. 
Therefore, the FR construction leads to an extended Poisson algebra that serves as a phase space for a classical model of $n$ interacting particles (and FR constructions for different choices of the classical $r$-matrix are physically inequivalent).

 In more technical terms, such PL structures on the isometry groups are defined in terms of classical $r$-matrices that have to fulfil a very specific compatibility condition  in order for them to be suitable to define the extended phase space: their symmetric component  has to be dual to the Ad-invariant symmetric bilinear form in the CS action \cite{AMII,FR,MScqg, MSquat,MS}. 
This condition gives rise to a strong constraint on the (initially quite large) zoo of possible $r$-matrices that can be defined on the Lie algebras of the isometry groups of (2+1) spacetimes,  and the quantisation of their associated PL structures will give rise to quantum group symmetries. Nonetheless, a sufficient condition for a given $r$-matrix in order to fulfil the above-mentioned FR condition~\cite{FR}  is provided by the fact that such $r$-matrix arise as the canonical $r$-matrix associated with a classical Drinfel'd double structure \cite{Drib, Dr}  of the isometry Lie algebra (see \cite{BHMcqg}). This way, Drinfel'd doubles (DD) neatly enter the scenario of (2+1) gravity, and a systematic study of all possible DD structures underlying Lorentzian gravity (with and without cosmological constant) has been recently undertaken in~\cite{BHMcqg}. Afterwards, the full quantum deformation of each of the two different families of Lorentzian classical $r$-matrices arising in this classification has been performed in \cite{BHMplb} and \cite{Naranjo}.

 It is also worth mentioning that in this context noncommutative spacetimes appear in a natural way already at the classical level, since the above-mentioned PL structure on the isometry group given by the FR-compatible $r$-matrix yields non-vanishing Poisson brackets among the local group parameters corresponding to time and space coordinates. This can be understood in the dual PL structure as a consequence of the fact that the momenta have a non-Abelian addition law, which is tantamount to the classical momentum space in (2+1) gravity being curved (see~\cite{Stalk} for a detailed account of this topic). Whereas these complementary features (noncommutativity of the spacetime coordinates---often related to the discreteness of quantum spacetime---and non-Abelian momentum space) appear in different approaches to (2+1) quantum gravity  (see~\cite{Snyder, Hooft, Wcqg,MWnc} and references therein), in the present context \emph{both} arise as properties linked to Hopf-algebraic properties of the PL structures defined on the gauge group.
Moreover, after quantisation, the PL structure on the isometry group (on its dual) is converted into a quantum group (into a quantum universal enveloping algebra) where the spacetime operators are non-commuting, since they come from the quantisation of the classical spacetime group coordinates that feature non-vanishing PL brackets. As a consequence, in the construction of the Hilbert space, its symmetries and observables, the quantum group associated with the FR classical $r$-matrix will play a prominent role.

The CS approach to non-relativistic quantum gravity in (2+1) dimensions had already been developed in \cite{Bernd1,Bernd2}. 
As it was pointed out 
in \cite{Bernd1}, (2+1)-Poincar\'e gravity as a CS gauge theory does not possess a well-defined non-relativistic limit. However, by considering a non-trivial two-fold central extension of the Galilei Lie algebra \cite{BLL,levy}, such a CS formulation is possible, and its non-vanishing cosmological constant counterpart was further constructed in \cite{Bernd2} by making use of the corresponding non-trivial two-fold central extension of the Newton-Hooke Lie algebra \cite{BLL}.
In both cases,  the DD  structure of these centrally extended (2+1) kinematical Lie algebras was explictly outlined, as well as its role in the construction of the corresponding quantisation. 

 Therefore, the relevant quantum groups for Galilean quantum gravity are the quantum doubles arising from the DD structures providing the FR classical $r$-matrix.
In the Galilei case~\cite{Bernd1}, the quantum double is a two-fold centrally extended Galilei-Hopf algebra with non-deformed commutation rules, 
which implies that the representations of the Galilei quantum double are just the classical ones. Nevertheless, the corresponding quantum $R$-matrix, whose first-order is the classical $r$-matrix, is non-trivial and provides the action of the generators of the braid group on the physical Hilbert space, and such an action is related to the scattering of two massive Galilean particles. On the other hand, the non-vanishing cosmological constant construction gives rise to the Newton-Hooke Lie algebra (as a double) discussed in~\cite{Bernd2}, where some properties of the quantum double, its associated noncommutative spacetimes and quantum $R$-matrix are sketched.

It is worth stressing that, from now on and for the sake of simplicity, we will use the generic adjective `Lorentzian'\ for the whole set of (anti)-de Sitter and Poincar\'e cases, and `Galilean'\ for  the non-relativistic set containing the Newton-Hooke and Galilei Lie algebras. When needed, each particular case (that correspond to different values of the cosmological constant $\Lambda$) will be explicitly identified.

The purpose of this article is two-fold. Firstly, by following the approach used in \cite{BHMplb,Naranjo}, we will construct the full quantum group (quantum DD) corresponding to the Galilean doubles introduced in \cite{Bernd1,Bernd2}  along with the associated noncommutative spacetimes corresponding to such quantisation. 
It is important to emphasise that the so obtained quantum non-relativistic spacetimes will have an `absolute time' coordinate in both the Galilei and Newton-Hooke cases.
Secondly, we will show that the DD and $r$-matrices found in \cite{Bernd1,Bernd2} as the keystones of the CS approach to Galilean (2+1) quantum gravity can be obtained as the appropriate non-relativistic limit of one specific family, among the various Lorentzian DD that have been found in \cite{BHMcqg}. For the sake of completeness, we will also present the results of the non-relativistic limit of all the remaining DD structures in \cite{BHMcqg}.

The paper is organised as follows. In section 2, the one-parameter family AdS$_{\omega}$ of Lie algebras of isometries of (2+1)-Lorenztian gravity ($\omega=-\Lambda)$, along with its 
non-relativistic version $\mathfrak{nh}_{\omega}(2+1)$, are described. Next, in section 3, we move on to present the two-fold central extensions of the AdS$_{\omega}$ and $\mathfrak{nh}_{\omega}(2+1)$ algebras, and we show that the latter can be obtained as an appropriate non-relativistic limit of the former by performing the appropriate Lie algebra contraction. 
Then, in section 4, after briefly reviewing the basics of DD, canonical $r$-matrices and quantum deformations, we construct the classical doubles underlying the two-fold centrally extended
(2+1)-Galilean Lie algebras of~\cite{Bernd1,Bernd2}, although in this case we will always work in a kinematical basis similar to the one used in the classification of (2+1)-Lorentzian DD  performed in \cite{BHMcqg}. In particular, we will firstly construct the DD corresponding to the vanishing cosmological constant case (the two-fold centrally extended (2+1)-Galilei Lie algebra) and, afterwards, we will carry out the same analysis for the non-vanishing cosmological constant setting (the two-fold centrally extended (2+1)-Newton-Hooke Lie algebra). 

Section 5 is devoted to the explicit 
construction of the full quantum group associated with (2+1)-Galilean gravity, {\em i.e.}, the quantum deformation of the NH$_{\omega}$ PL group (associated with $\mathfrak{nh}_{\omega}(2+1)$) that is generated from the canonical $r$-matrix coming from the Galilean doubles described in section 4. In particular, the noncommutative spacetime (with cosmological constant) that arises in this quantum double is given by
\be
\left[\hat{x}_1,\hat{x}_2\right]=f_\omega\left(\hat{x}_0\right),
\qquad
\left[\hat{x}_1,\hat{x}_0\right]=\left[\hat{x}_2,\hat{x}_0\right]=0,
\label{ncintro}
\ee
where $f_\omega\left(\hat{x}_0\right)$ is a formal power series of the `time coordinate' $\hat{x}_0$ that becomes linear in the vanishing cosmological constant limit $\omega\to 0$. Since $\hat{x}_0$ commutes with all the remaining quantum group coordinates, it can thus be interpreted as an `absolute quantum time' coordinate, as it should be expected in a non-relativistic setting. However, it is worth stressing that quantum gravity effects turn out to be non-trivial for the quantum space coordinates $(\hat{x}_1,\hat{x}_2)$, that become noncommutative ones.

Next, section 6 is devoted to perform a complete analysis of the non-relativistic limit of all the DD Lorentzian $r$-matrices found in \cite{BHMcqg} for the AdS$_{\omega}$ algebra. As a result, we show that the Galilean canonical $r$-matrix studied here and in \cite{Bernd1,Bernd2} is exactly the non-relativistic limit of the AdS$_{\omega}$ $r$-matrix whose quantum group has recently been constructed in \cite{BHMplb}. Moreover, it is shown how the noncommutative Galilean spacetime~\eqref{ncintro} can also be obtained as the appropriate non-relativistic contraction $c\to \infty$ of the AdS$_{\omega}$ noncommutative spacetime with cosmological constant presented in \cite{BHMplb}. Surprisingly enough, it is also found that the non-relativistic limit of the other family of AdS$_{\omega}$ $r$-matrices studied in \cite{BHMplb} (which is just the one related to so-called $\kappa$-deformations, as it was shown in \cite{Naranjo}) always yields trivial (commutative) quantum spacetimes in the Galilean limit. Finally, section 7 closes the paper with the 
main conclusions as well as with some open problems.


\sect{(2+1)-Galilean Lie algebras and spacetimes}

Let us consider the  real  Lie algebras of the  isometry groups of  the (2+1)-dimensional (anti)-de Sitter and Minkowski spaces
    denoted, in a unified setting,   by   $\mbox{AdS}_\omega\simeq\mathfrak{so}_{\k}(2,2)$, where $\k$ is a  real contraction parameter that is  directly related to the cosmological constant through $\omega=-\Lambda$.
Let $\{P_0, P_i, K_i, J\}$ $(i=1,2)$  be  the infinitesimal generators of  a time translation, spatial translations,  boosts and a spatial rotation, respectively. In such a basis,  the Lie brackets defining AdS$_\omega$  are given by
\begin{align}
[J,K_i] &=\epsilon_{ij}K_j\,, & [K_1,K_2] &=-\frac {1}{c^2}\,J\,, & [K_i,P_0] &=P_i\,, \nonumber  \\[8pt] 
\label{ba} 
[J,P_i] &=\epsilon_{ij}P_j\,, & [P_1,P_2] &=-\frac{\k}{c^2}\,J\,, & [P_0,P_i] &=\k K_i\,, \\[6pt]
[J,P_0] &=0\,, & [K_i,P_j] &=\frac{1}{c^2}\,\delta_{ij}P_0\,, \nonumber
\end{align}
where $\epsilon_{ij}$ is a skew-symmetric tensor   such that $\epsilon_{12}=1$  and $c$ is the {speed of light}, that we manifestly write in order to make the effect of the non-relativistic limit $c\to\infty$ more evident.
The two (quadratic) Casimir
invariants of the Lie algebra $\mbox{AdS}_\omega$ read
\begin{align}
{\cal C} &=- \frac {1}{c^2}\, P_0^2+\boldsymbol{\mathrm{P}}^2+\k\left(\boldsymbol{\mathrm{K}}^2-\frac{1}{c^2}\,J^2\right)\,, \nonumber \\[8pt] 
\label{bb}
 {\cal W}& =-\frac {1}{c^2}\, J P_0 +K_1 P_2  -K_2 P_1\,,
\end{align}
which correspond, in this order,  to the energy and
angular momentum of a particle in the free kinematics of the underlying relativistic spacetime. Hereafter, we denote $\>P=(P_1,P_2)$,  $\>P^2=P_1^2+P_2^2$ and similarly for any other two-dimensional vector quantity.

The  (2+1) AdS, dS and Minkowski spaces are obtained as symmetric homogeneous
Lo\-rentzian spacetimes  with  constant  curvature  given by the quotients 
$
 {\rm SO}_{\k}(2,2)  /{\rm  SO}(2,1)  
$,
 where    ${\rm SO}_{\k}(2,2)$ is the Lie group of  AdS$_\omega$ and ${\rm  SO}(2,1)$ is the Lorentz isotopy subgroup spanned by $\langle J,\>K\rangle $.
 In this approach, the parameter $\k$ is just the {constant sectional curvature} of the spacetime which, in turn, 
  can be written in terms of the (time) {universe
radius} $\tau$ and the {cosmological constant}  $\Lambda$ as
\be
\k=\pm 1/\tau^2=-\Lambda .
\label{bc}
\ee
Hence, according to a positive/zero/negative value of $\k$,  we recover the AdS/Minkowski/dS spacetimes and their corresponding isometry Lie  groups, which are displayed in the first column of Table~\ref{table1}.

For our further purposes, let us parametrise these spaces through so-called {\em geodesic
parallel coordinates}  $(x_0,x_1,x_2)$ \cite{conformes}, which can be regarded as a generalisation of the flat  Cartesian coordinates to the non-vanishing curvature case.
Essentially, $x_0$ is defined as the geodesic distance along a time-like geodesic, whereas $\>x$ are geodesic distances along   space-like ones
(for more details, see~\cite{Naranjo}). In terms of these coordinates, the metric on the three (2+1)-dimensional Lorentzian spacetimes turns out to be
\be
\dd\sigma^2 = \cosh^2(\sqrt{\k} \,x_2/c)\cosh^2(\sqrt{\k} \,x_1/c)\, \dd x_0^2-\frac 1 {c^2}\cosh^2(\sqrt{\k} \,x_2/c)\, \dd x_1^2-\frac 1 {c^2}\,\dd x_2^2 .
\label{metric1}
\ee

 Now it becomes evident how the  explicit presence of the speed of light $c$  in the structure constants of the commutation relations (\ref{ba})  and in the Casimir invariants (\ref{bb}) of $\mbox{AdS}_\omega$ allows one to perform a well-defined {\em non-relativistic limit}  which  straightforwardly  leads to the corresponding Galilean Lie algebras $\mathfrak{nh}_\k(2+1)$ and Casimirs in the form
\begin{align}
[J,K_i] &=\epsilon_{ij}K_j\,, & [K_1,K_2] &=0\,, & [K_i,P_0] &=P_i\,, & [K_i,P_j] &=0\,, \nonumber \\[8pt] 
\label{bd} [J,P_i] &=\epsilon_{ij}P_j\,, & [P_1,P_2] &= 0\,, & [P_0,P_i] &=\k K_i\,, &  [J,P_0] &=0\,, 
\end{align}
\begin{equation}
\begin{array}{l}
\displaystyle{ {\cal C}=  \>P^2+\k  \>K^2  
,}\qquad 
\displaystyle{  {\cal W}= K_1 P_2  -K_2 P_1  }.
\end{array}
\label{be}
\end{equation}
In this   non-relativistic case, the   three    (2+1)-dimensional     symmetric homogeneous spacetimes  with  constant   curvature~\cite{BHOS, BLL,expansions,brazil}   are  the two  Newton-Hooke (NH) spacetimes plus the Galilean one. All of them can be  simultaneously defined as the quotient
\be
 \mathrm{NH}_{\k}(2+1) /{\rm  ISO}(2)  ,\quad\mbox{with}\quad  {\rm  ISO}(2)\simeq \RR_2\odot {\rm SO}(2),\qquad \RR_2=\langle \>K\rangle,\qquad {\rm SO}(2)=\langle J\rangle,
\nonumber
\ee
where  $ \mathrm{NH}_{\k}(2+1)$ denotes the Lie group with Lie algebra given by $\mathfrak{nh}_\k(2+1)$,  `$\odot$' is a semidirect  product and ${\rm  ISO}(2) $ is the Euclidean isotopy subgroup  (which comes from the non-relativistic limit of the Lorentz group ${\rm  SO}(2,1)$).
Explicitly:

\begin{itemize}

\item  When $\k=1/\tau^2>0$ we obtain the {\em oscillating} NH spacetime with the following structure for its isometry group~\cite{expansions,Wolf}:
   $$
    {\bf NH}_+^{2+1}={\rm T}_4\left( {\rm SO(2)} \oplus  {\rm SO(2)}   \right) /{\rm ISO}(2),\quad {\rm T}_4=\langle \>P,\>K\rangle,\quad {\rm SO(2)}=\langle J\rangle,\quad {\rm SO(2)}=\langle P_0\rangle .
   $$
  
 \item When $\k=-1/\tau^2<0$ we recover the  {\em  expanding} NH spacetime as
   $$
    {\bf NH}_-^{2+1}={\rm T}_4\left( {\rm SO(2)} \oplus  {\rm SO(1,1)}   \right) /{\rm ISO}(2),\  {\rm T}_4=\langle \>P,\>K\rangle,\  {\rm SO(2)}=\langle J\rangle,\  {\rm SO(1,1)}=\langle P_0\rangle .
   $$

\item Finally, the  flat limit  $\k=0$ ($\tau\to \infty$)  leads to the Galilean spacetime
   \bea
 &&   {\bf G}^{2+1}= {\rm IISO(2)}  /{\rm ISO}(2),\quad
     {\rm IISO(2)} \simeq   \RR_3\odot\left( \RR_2\odot {\rm SO}(2)\right),\nonumber\\[2pt]
     && \RR_3=\langle P_0, \>P\rangle,\qquad \RR_2=\langle \>K\rangle,\qquad {\rm SO}(2)=\langle J\rangle.
     \nonumber
 \eea
 
 \end{itemize}

The non-relativistic limit of the metric (\ref{metric1}) provides a degenerate time-like
metric,  $\dd\sigma^2 = \dd x_0^2$, 
which, in fact, corresponds to an `absolute-time', such that the Galilean  spacetimes  have foliations with leaves defined by a constant time $x_0= t_0$.
Nevertheless, the complete metric structure of these three spaces can be obtained through a non-degenerate `subsidiary' (two-space Euclidean) metric $\dd\sigma^2_{(2)}$ that is obtained as $\lim_{c\to \infty}  -c^2\dd\sigma^2$ and restricted to $x_0= t_0$ (see~\cite{conformes} for details), namely,
\be
\dd\sigma^2_{(2)} = \dd x_1^2+\dd x_2^2 ,\qquad x_0= t_0.
\label{metric2}
\ee
Therefore, these spaces are characterised by an invariant foliation whose leaves are 
the `absolute space' at the time $x_0= t_0$, endowed  with a non-degenerate subsidiary spatial metric  (\ref{metric2}). Recall that differences among these three 
Galilean  spacetimes arise when the connection is considered~\cite{conformes}. All the above results are summarised in Table~\ref{table1}.

\begin{table}[t]
{\footnotesize
 \noindent
\caption{{The (2+1)-dimensional Lorentzian spacetimes with constant curvature $\k=\pm 1/\tau^2$  and their  corresponding Galilean spacetimes ($c\to \infty$), together with their metrics in geodesic parallel coordinates.}}
\label{table1}
\medskip
\noindent\hfill
$$
\begin{array}{ll}
\hline
\\[-6pt]
{\mbox{  
 Lorentzian spacetimes}}  &\quad{\mbox  { 
 Galilean spacetimes}}\\[4pt] 
\hline
\\[-6pt]
\mbox {$\bullet$ AdS:}\ \k=1/\tau^2>0,\  \Lambda<0&\quad\mbox {$\bullet$ Oscillating NH:}\ \k=1/\tau^2>0,\  \Lambda<0\\[4pt]
 \displaystyle{{   {\bf AdS}^{2+1}={\rm SO}(2,2)/{\rm SO}(2,1) } }&\quad
 \displaystyle{{    {\bf NH}_+^{2+1}={\rm T}_4\left( {\rm SO(2)} \oplus  {\rm SO(2)}   \right) /{\rm ISO}(2) }} \\[4pt]
 \dd\sigma^2 = \cosh^2(\frac{ \,x_2}{c\tau })\cosh^2(\frac{ \,x_1}{c\tau })\dd x_0^2-\frac 1 {c^2}\cosh^2(\frac{ \,x_2}{c\tau })\dd x_1^2-\frac 1 {c^2}\,\dd x_2^2
  &\quad
\dd\sigma^2 = \dd x_0^2
\\[4pt]
  \dd\sigma^2_{(2)}= -c^2\dd\sigma^2 &\quad
\dd\sigma^2_{(2)} = \dd x_1^2+\dd x_2^2 ,\qquad x_0= t_0
  \\[8pt]
\mbox {$\bullet$ Minkowski:}\ \k=\Lambda=0,\ \tau\to \infty&\quad\mbox {$\bullet$ Galilei:}\ \k=\Lambda=0,\ \tau\to \infty\\[4pt]
  \displaystyle{      {\bf
M}^{2+1}={\rm ISO}(2,1)/{\rm SO}(2,1) }&\quad
 \displaystyle{       {\bf
G}^{2+1}={\rm IISO(2)}/{\rm ISO(2)} } \\[4pt]
 \dd\sigma^2 = \dd x_0^2-\frac 1 {c^2}\,\dd x_1^2-\frac 1 {c^2}\,\dd x_2^2
  &\quad
\dd\sigma^2 = \dd x_0^2
\\[4pt]
  \dd\sigma^2_{(2)}= -c^2\dd\sigma^2 &\quad
\dd\sigma^2_{(2)} = \dd x_1^2+\dd x_2^2 ,\qquad x_0= t_0
 \\[8pt]
\mbox {$\bullet$ dS:}\ \k=-1/\tau^2<0,\  \Lambda>0&\quad\mbox {$\bullet$ Expanding NH:}\ \k=-1/\tau^2<0,\  \Lambda>0\\[4pt]
\displaystyle{{   {\bf dS}^{2+1}={\rm SO}(3,1)/{\rm SO}(2,1) } }&\quad
 \displaystyle{{    {\bf NH}_-^{2+1}={\rm T}_4\left( {\rm SO(2)} \oplus  {\rm SO(1,1)}   \right)  /{\rm ISO}(2) }} \\[4pt]
  \dd\sigma^2 = \cos^2(\frac{ \,x_2}{c\tau })\cos^2(\frac{ \,x_1}{c\tau })\dd x_0^2-\frac 1 {c^2}\cos^2(\frac{ \,x_2}{c\tau })\dd x_1^2-\frac 1 {c^2}\,\dd x_2^2
  &\quad
\dd\sigma^2 = \dd x_0^2
\\[4pt]
  \dd\sigma^2_{(2)}= -c^2\dd\sigma^2 &\quad
\dd\sigma^2_{(2)} = \dd x_1^2+\dd x_2^2 ,\qquad x_0= t_0
\\[8pt]
\hline
\end{array}
$$
\hfill}
\end{table}


\sect{Centrally extended (2+1)-Galilean Lie algebras}

It is well-known that the second cohomology group of the (2+1)-NH Lie groups is two-dimensional~\cite{azca}, and the same is true for the (2+1)-Galilean one~\cite{levy,gadella}. This means that   NH$_{\k}$ can be endowed with a maximum of two non-trivial {\em central extensions} associated with two additional central generators,  hereafter denoted by $\mm$ and $\sss$.

Such a two-fold central extension of the (2+1)-Galilean Lie algebras can be recovered through a `pseudoextension' approach~\cite{azca,azca2, azca3} from the 
Lorentzian groups  ${\rm SO}_{\k}(2,2)$. 
In particular, let us consider the extended Lie algebra AdS$_\omega\oplus \RR_2$ and the following   redefinition for the   generators:
 \be
 P'_0= P_0-c^2\mm\,, \qquad P'_i=P_i\,, \qquad K'_i=K_i\,,  \qquad J'=J+c^2 \sss\,, \qquad i=1,2\,.
 \label{ca}
 \ee
By introducing this map into the commutation relations (\ref{ba}) and Casimirs (\ref{bb}) we obtain 
\begin{align}
[J',K'_i] &=\epsilon_{ij}K'_j\,, & [K'_1,K'_2] &=-\frac {1}{c^2}\,J'+\sss\,, & [K'_i,P'_0] &=P'_i\,, \nonumber \\[8pt] 
\label{cb} 
[J',P'_i] &=\epsilon_{ij}P'_j\,, & [P'_1,P'_2] &=-{\k} \left(\frac{1}{c^2}\,J' - S \right)\,, & [P'_0,P'_i] &=\k K'_i\,, \\[8pt]
[J',P'_0] &=0\,, & [K'_i,P'_j] &=\delta_{ij}\left( \frac{1}{c^2}\,P'_0 +\mm\right)\,, & [\mm,\,\cdot\,] &=[\sss,\,\cdot\,]=0\,, \nonumber 
\end{align}
\begin{align}
 {\cal C} &=- \frac {1}{c^2}\left(  {P'}_0^2 +c^4 \mm^2 +2 c^2 P'_0\mm \right) +\>{P'}^2+\k \>{K'}^2 \nonumber \\[8pt]
\label{cc}
 & \,\,\,\,\,\,\, {}  -\frac{\k}{c^2}   \left (  {J'}^2 +c^4\sss^2-2 c^2 J'\sss \right)\,, \\[8pt] 
{\cal W} &=-\frac 1{c^2}\left( J' P'_0 -c^4 \mm\sss+c^2J'\mm-c^2P'_0\sss\right)+K'_1 P'_2  -K'_2 P'_1\,. \nonumber
\end{align}

Obviously, when $c$ is a finite number  both central extensions $\mm$ and $\sss$ are trivial ones, since the Lie algebra can be decoupled as AdS$_\omega\oplus \RR_2$ through the inverse of the transformation (\ref{ca}) (recall that
the second cohomology group of any semisimple group is always trivial). Nevertheless, when the non-relativistic contraction $c\to \infty$ is applied to (\ref{cb}), the former pseudoextensions become non-trivial ones, and we obtain the commutation relations of the two-fold centrally extended Galilean algebras  $\overline {\mathfrak{nh}_{\k}(2+1)}$:
\begin{equation}
\begin{array}{llll}
\displaystyle{ [J,K_i]=\epsilon_{ij}K_j},&\quad \displaystyle{ [K_1,K_2]= 
 \sss},&\quad
\displaystyle{ [K_i,P_0]=P_i},&\quad  \\[8pt] 
\displaystyle{ [J,P_i]=\epsilon_{ij}P_j },&\quad \displaystyle{[P_1,P_2]= 
{\k}    S  },&\quad
\displaystyle{[P_0,P_i]=\k K_i},&\quad   \\[8pt]
\displaystyle{   [J,P_0]=0    }  ,&\quad \displaystyle{[K_i,P_j]=\delta_{ij}   
 \mm  },&\quad [\mm,\,\cdot\,]=[\sss,\,\cdot\,]=0 ,
\end{array}
\label{cd}
\end{equation}
where,  for the sake of simplicity,  we have dropped the `tilde' in the contracted generators. As   can be checked,  in this algebra the central generators cannot be decoupled by any redefinition of the basis, and this Lie algebra is no longer a direct sum with respect to $\RR_2$.

The corresponding Casimirs for  $\overline {\mathfrak{nh}_{\k}(2+1})$ (besides $\mm$ and $\sss$, which are also central generators) arise from (\ref{cc})  by considering the limits
\begin{align}
C &=\lim_{c\to\infty}\left( {\cal C} +c^2 \mm^2  + {\k}{c^2} \sss^2\right)\,, & W &=\lim_{c\to\infty}\left( {\cal W} -c^2 \mm\sss\right)\,, \nonumber
\end{align}
in which the additional constant terms depending on the central extensions and on the speed of light have to be considered in order to guarantee the convergence of the limit. The result is  (again, we have suppressed the tilde)
\begin{align}
C &=\>{P}^2+\k \>{K}^2  - 2   P_0\mm+2\k J \sss\,, & W &=  K_1 P_2  -K_2 P_1 -   J\mm+P_0\sss\,. \nonumber
\end{align}
As expected, when the central extensions vanish the commutation relations for $\mathfrak{nh}_\k(2+1)$ (\ref{bd}) and its Casimirs (\ref{be}) are straightforwardly  recovered.

Notice that the physical meaning of the central generators can roughly be inferred from the map (\ref{ca}): 
 $\mm$ can be interpreted as a {mass}, while $\sss$ can be understood as an angular momentum/spin. 
In fact, the `exotic' central extension $\sss$ of the (2+1)-Galilei Lie algebra was shown in~\cite{gadella} to generate  a  non-relativistic remainder of the Thomas precession, and its significance for (2+1)-dimensional  classical and quantum physics has extensively been discussed in the literature (see \cite{LukiStichel,Jackiw,DuvalPLB,Horvathy} and references therein). On the other hand, the two-fold extended NH  Lie algebras have also been studied in, for instance, \cite{Alvarez,Olmo}, and the cosmological significance of its associated spacetimes in the (3+1)-dimensional case was discussed in \cite{Gibbons}.


\sect{Drinfel'd doubles for (2+1)-Galilean gravity}

Having motivated and discussed in some detail the three (2+1)-Galilean spacetimes, let us turn our analysis to the explicit construction of the corresponding DD structures underlying the two-fold centrally extended Galilei and NH  (2+1) Lie algebras (we recall that the former is simply the limit of vanishing cosmological constant of the latter). To this  aim, we will make use of the approach followed in~\cite{BHMplb} for the study of one of the DD  structures of the AdS$_\omega$ algebra.

\subsection{Drinfel'd doubles, Lie bialgebras and quantum deformations}

A $2d$-dimensional Lie algebra $\mathfrak{a}$ has the  structure of a (classical) DD~\cite{Dr}   if there exists a basis $\{X_1,\dots,X_d,x^1,\dots,x^d \}$ of $\mathfrak a$ in which the 
Lie bracket takes the form
\begin{align}
[X_i,X_j]= c^k_{ij}X_k\,, \qquad  
[x^i,x^j]= f^{ij}_k x^k\,, \qquad
[x^i,X_j]= c^i_{jk}x^k- f^{ik}_j X_k\,.
\label{agd}
\end{align}
This definition means that the two sets of generators $\{X_1,\dots,X_d\}$ and $\{x^1,\dots,x^d \}$ form two Lie subalgebras with structure constants  $c^k_{ij}$ and $f^{ij}_k$, respectively.  Moreover,  the expression for the crossed brackets $[x^i,X_j]$ implies  that there exists an  Ad-invariant symmetric bilinear form on $\mathfrak{a}$ given by
\begin{align}\label{pairdd}
 \langle X_i,X_j\rangle= 0\,, \qquad \langle x^i,x^j\rangle=0\,, \qquad
\langle x^i,X_j\rangle= \delta^i_j\,, \qquad \forall i,j\,.
\end{align}
Furthermore,  it is readily proven from~\eqref{agd} that a quadratic Casimir operator for $\mathfrak{a}$ is given by
\begin{align}
C=\tfrac12\sum_i{(x^i\,X_i+X_i\,x^i)}. 
\nonumber
\end{align}

Next, DD are in one-to-one correspondence with Lie bialgebra structures, and play an essential role in quantum group theory (see~\cite{CP,majid} for details). We recall that, given a Lie algebra $\mathfrak{g}$ in a certain basis $\{X_i\}$, a
 Lie bialgebra structure ($\mathfrak{g},\delta$) is given by
 \be
[X_i,X_j]= c^k_{ij}X_k\,, \qquad  \delta(X_n)=f_{n}^{lm} X_l\wedge X_m\,,
\nonumber
\ee
where $\delta:\mathfrak{g}\rightarrow \mathfrak{g}\wedge \mathfrak{g}$ is the so-called cocommutator map,
whose cocycle condition takes the form of a  compatibility condition between the
structure tensors $c$ and $f$, namely
\be
f^{ab}_k c^k_{ij} = f^{ak}_i c^b_{kj}+f^{kb}_i c^a_{kj}
+f^{ak}_j c^b_{ik} +f^{kb}_j c^a_{ik}  \;. \label{agb}
\ee  
Denoting by  $\{x^i\}$ the  basis of   $\mathfrak{g} ^*$ 
dual to $\{X_i\}$, we can consider an associated `double' vector space 
$\mathfrak{a}=\mathfrak{g} \oplus \mathfrak{g}^*$, where the bilinear form~\eqref{pairdd} provides a canonical pairing between $\mathfrak{g}$ and $\mathfrak{g} ^*$, 
and $\mathfrak{a}$ can be endowed with a Lie algebra
structure by means of the brackets (\ref{agd}). In fact, the Jacobi identity for this bracket is nothing but the compatibility condition~\eqref{agb}. Consequently, 
the resulting  double-dimensional Lie algebra $\mathfrak{a}\equiv D ({\mathfrak{g}})$ is called the
 DD Lie algebra of $(\mathfrak{g},\delta)$, and (the connected component of) its Lie group  is the DD Lie group associated with $(\mathfrak{g},\delta)$. 
 
Furthermore, if $\mathfrak a$ is a DD Lie algebra, its corresponding Lie group  can always be endowed with a PL structure generated by the canonical  classical
$r$-matrix
\be
r=\sum_i{x^i\otimes X_i} 
\label{canonr}
\ee
which is a (constant) solution of the classical Yang-Baxter equation
$[[r,r]]=0.
$
The skew-symmetric component of such $r$-matrix is
\be  r'=\tfrac12 \sum_i{x^i\wedge X_i}  ,
\label{rmat}
\ee
and the symmetric component $\Omega$ coincides with the tensorised form of the canonical quadratic Casimir element in $\mathfrak a$
\be\label{omega}
\Omega=r-r'=\tfrac12\sum_i{(x^i\otimes X_i+X_i\otimes x^i)},
\nonumber
\ee
which is just the Fock-Rosly condition for a PL structure in order to be compatible with the requirements of a CS (2+1) gravity theory with gauge group given by the DD Lie group associated with $\mathfrak a$. This also implies that $\mathfrak{a}\equiv D ({\mathfrak{g}})$  can be endowed with a
(quasi-triangular) Lie bialgebra structure $(D ({\mathfrak{g}}),\delta_{D})$ that is determined by the canonical  classical
$r$-matrix~\eqref{rmat} through
\be
\delta_{D}(Y)=[ Y \otimes 1+1\otimes Y ,  r],
\quad
\forall Y\in D ({\mathfrak{g}}).
\label{rcanon2}
\ee

Finally, DD Lie algebras and quantum deformations are directly connected, since each quantum universal enveloping algebra  $(U_z(\mathfrak g),\Delta_z)$ of a Lie algebra  $\mathfrak g$  is associated with  a
unique Lie bialgebra structure
$(\mathfrak g,\delta)$.   In fact, the cocommutator $\delta$ is just the skew-symmetric part of the first-order expansion of the coproduct  $\Delta_z$ in terms of the deformation parameter $z$:
\be
\delta (X)=\frac 1 2\big(\Delta_z (X)-\sigma\circ \Delta_z (X)\big) +\mathcal O\left(z^2\right)\,, \qquad \forall X\in \mathfrak g\,, \nonumber
\ee
where $ \sigma$ is the flip operator  $ \sigma (X\otimes
Y)=Y\otimes X$. Therefore, if a Lie algebra $\mathfrak{a}$ has a DD structure (\ref{agd}),  this implies that  $(\mathfrak{a},\delta_{D})$ is a Lie bialgebra given by~\eqref{rcanon2} through the canonical $r$-matrix given by (\ref{canonr}). Therefore, there exists a quantum universal enveloping algebra $(U_z(\mathfrak{a}),\Delta_z)$ whose first-order coproduct is given by $\delta_{D}$, and this quantum deformation can be viewed as  the quantum symmetry corresponding to the given DD structure for $\mathfrak{a}$. Therefore, if the CS theory on the model spacetime with isometries given by a DD Lie algebra $\mathfrak{a}$ is considered, the $r$-matrix (\ref{canonr}) will be compatible with the Fock-Rosly condition and the quantum group constructed as the (Hopf algebra) dual to the quantum algebra  $(U_z(\mathfrak{a}),\Delta_z)$ will be expected to provide the noncommutative geometry underlying the combinatorial quantisation of the theory.

In the sequel, we will explicitly show that the centrally extended (2+1) Galilei and NH  algebras are DD Lie algebras $\mathfrak{a}\equiv D ({\mathfrak{g}})$, and this double structure will induce a canonical quantum deformation for both of them, which will be fully constructed and analysed in the remaining sections of the paper, and that will give rise to a distinguished noncommutative and non-relativistic (2+1)-dimensional spacetime.


\subsection{Centrally extended (2+1)-Galilei Lie algebra as a Drinfel'd double}

Following \cite{Bernd1}, let us consider the centrally extended two-dimensional Euclidean Lie algebra, $\overline{\mathfrak{e}(2)}$, defined by the following Lie brackets:
\begin{align}
[X_1,X_2]& =X_4\,, & [X_1,X_3]&=-X_2\,, & [X_2,X_3]&=X_1\,, & [X_4,\, \cdot\,]&=0\, , \label{e2e}
\end{align}
where $X_4$ is the generator corresponding to the central extension, $X_3$ is the rotation generator and $X_1,X_2$ generate the two translations on the plane.
As mentioned in \cite{Bernd1}, this Lie algebra is sometimes called the Nappi-Witten Lie algebra \cite{NappiWitten}. The trivial Lie bialgebra structure  $\left(\overline{\mathfrak{e}(2)},\delta_{0}\right)$ is given by the cocommutator
\begin{equation}
 \delta_0 (X_i)=0\,, \qquad \forall i\,, \nonumber
\end{equation}
which corresponds to the undeformed coproduct for the universal enveloping algebra $U\Big(\overline{\mathfrak{e}(2)}\Big)$:
\begin{equation}
\Delta_0 (X_i)=X_i\otimes 1+1\otimes X_i\,, \qquad \forall i\,. \nonumber
\end{equation}

The DD Lie algebra $D \Big(\overline{\mathfrak{e}(2)}\Big)$ associated with the Lie bialgebra $\Big(\overline{\mathfrak{e}(2)},\delta_{0}\Big)$ is straightforwardly obtained by considering the canonical pairing
\begin{align}
\langle X_i,X_j\rangle &=0\,, & \langle x^i,x^j\rangle &=0\,, & \langle x^i,X_j\rangle &=\delta _{ij}\,, \nonumber 
\end{align}
together with the remaining DD Lie brackets~\eqref{agd}, which read:
\begin{align}
\left[x^i,x^j\right]&=0\,, & \forall i,j&=1,\dots,4\,, \nonumber \\
\left[x^1,X_1\right]&=0\,, & \left[x^1,X_2\right]&=x^3\,, & \left[x^1,X_3\right]&=-x^2\,, & \left[x^1,X_4\right]&=0\,,  \nonumber \\
\left[x^2,X_1\right]&=-x^3\,, & \left[x^2,X_2\right]&=0\,, & \left[x^2,X_3\right]&=x^1\,, & \left[x^2,X_4\right]&=0\,,  \nonumber \\
\left[x^3,X_1\right]&=0\,, & \left[x^3,X_2\right]&=0\,, & \left[x^3,X_3\right]&=0\,, & \left[x^3,X_4\right]&=0\,,  \nonumber \\
\left[x^4,X_1\right]&=x^2\,, & \left[x^4,X_2\right]&=-x^1\,, & \left[x^4,X_3\right]&=0\,, & \left[x^4,X_4\right]&=0\,.  \label{ddg}
\end{align}

Now, if we perform the following change of basis
\begin{align}
X_1 & \equiv K_1\,, & X_2 & \equiv K_2\,, & X_3 & \equiv J\,, & X_4 & \equiv S\,, \nonumber \\
x^1 & \equiv -P_2\,, & x^2 & \equiv P_1\,, & x^3 & \equiv M\,, & x^4 & \equiv -P_0\,, \nonumber  
\end{align}
we are led to the two-fold centrally extended (2+1)-Galilei Lie algebra, $\overline{\mathfrak{g}(2+1)}$, given by~\eqref{cd} with $\omega=0$, namely:
\begin{align}
[J,K_i]&=\epsilon_{ij}K_j\,, & [K_1,K_2]&=\sss\,, & [K_i,P_0]&=P_i\,,\nonumber \\[8pt] 
[J,P_i]&=\epsilon_{ij}P_j\,, & [P_1,P_2]&=0\,, & [P_0,P_i]&=0\,,\nonumber \\[8pt]
[J,P_0]&=0\,, & [K_i,P_j]&=\delta_{ij}\mm\,, & [\mm,\,\cdot\,]&=[\sss,\,\cdot\,]=0\,, \nonumber
\end{align}
where $K_i\, (i=1,2)$ and $P_a$ $(a=0,1,2)$ are, respectively, the generators of Galilean boosts transformations and spacetime translations. 

As a consequence of this DD structure, 
$\overline{\mathfrak{g}(2+1)}$ is automatically endowed with a quasi-triangular Lie bialgebra structure generated by the canonical classical $r$-matrix (\ref{canonr}), whose expression in terms of the Galilei generators becomes
\be
r=-P_2 \otimes K_1 +P_1\otimes K_2+M\otimes J-P_0\otimes S\,.\nonumber
\ee
Note that this $r$-matrix is just (up to a global sign) the one given in Eq.~(4.33) in~\cite{Bernd1}, provided that the following relation between the generators of $\overline{\mathfrak{g}(2+1)}$ in~\cite{Bernd1} and the basis employed here is considered:
$$
J\rightarrow - J,
\qquad
J_i\rightarrow\epsilon_{ij}\,K_j,
\qquad
P_i\rightarrow P_i.
$$
Indeed, this classical $r$-matrix defines the PL and symplectic structures needed in order to develop the FR approach to (2+1) gravity without cosmological constant (see~\cite{we2, Bernd1}).

As is well-known, the cocommutator map only depends on the skew-symmetric component of the classical $r$-matrix (\ref{rmat}), which yields 
\be
r^{\prime}=\tfrac{1}{2}\left(K_1\wedge P_2+P_1\wedge K_2+M\wedge J+S\wedge P_0\right)\,,\nonumber
\ee
which will generate a quantum deformation of $\overline{\mathfrak{g}(2+1)}$ consisting of a superposition of four Reshetikin twists (each of the four components of the $r$-matrix contains generators whose commutator vanishes). 

In order to construct and analyse this deformation, it is convenient to explicitly introduce the quantum deformation parameter $\xi$ as a global multiplicative factor for the classical $r$-matrix generating the deformation:
\be
r_{\xi}\equiv\xi\,r^{\prime}=\tfrac{1}{2}\xi\left(P_1\wedge K_2+K_1\wedge P_2+M\wedge J+S\wedge P_0\right)\,. \nonumber
\ee
The cocommutator associated with $r_{\xi}$ is readily obtained, namely 
\begin{align}
\delta _{\xi}(J)&=0\,, & \delta _{\xi}(K_1)&=0\,, & \delta _{\xi}(K_2)&=0\,, & \delta _{\xi}(S)&=0\,, \nonumber \\
\label{coco1} \delta _{\xi}(P_1)&=\xi P_2\wedge M\,, & \delta _{\xi}(P_2)&=\xi M\wedge P_1\,, & \delta _{\xi}(P_0)&=\xi P_2\wedge P_1\,, & \delta _{\xi}(M)&=0\,,  
\end{align}
which corresponds to the first-order quantum coproduct. Therefore, \eqref{coco1} means that the full quantum deformation 
$U_{\xi}\Big(D\big(\overline{\mathfrak{e}(2)}\big)\Big)\sim U_{\xi}\left(\overline{\mathfrak{g}(2+1)}\right)$ possesses a non-deformed rotation-boosts sector enlarged with the central extension $S$,
\be
\Delta _{\xi}(Y)=\Delta _0(Y)=Y\otimes 1+1\otimes Y\,, \qquad Y=J, K_i, S\,, \nonumber
\ee
thereby the full deformation occurring in the translation sector $P_a$, which closes an Abelian subalgebra before deformation. As a result, both the extended (by $S$) rotation-boosts sector and the extended (by $M$) translation sector are expected to span Hopf subalgebras after the full deformation is constructed.

On the other hand, by defining the noncommutative coordinate functions dual to the kinematical quantum algebra generators by means of the following pairing
\begin{align}
\langle\hat{\vartheta}_0,J \rangle &= 1\,, & \langle\hat{\vartheta}_i,K_j \rangle &=  \delta _{ij}\,,& \langle\hat{s},S \rangle &=  1\,, \nonumber \\
\label{pairing} \langle\hat{x}_a,P_b \rangle &= \delta _{ab}\,,& \langle\hat{m},M \rangle &=  1\,, 
\end{align}
where $i,j=1,2,\, a,b=0,1,2$, the first-order quantum group dual to $U_{\xi}\left(\overline{\mathfrak{g}(2+1)}\right)$ will be given by the only non-vanishing relations coming from dualising (\ref{coco1}), namely 
\begin{align}
[\hat{x}_1,\hat{x}_2]&=-\xi\hat{x}_0\,,    & [\hat{x}_0,\,\cdot\,]&=0\, , \nonumber \\
\label{spacetime1} 
[\hat{x}_1,\hat{m}]&=-\xi\hat{x}_2\,,  & [\hat{x}_2,\hat{m}]&=\xi\hat{x}_1\,.
\end{align}

Although the full construction of the quantum group will be presented in section 5, it is already worth noticing that the first-order relations (\ref{spacetime1}) are quite remarkable, since the scenario we are dealing with, the quantum Galilei group, carries an absolute time $\hat{x}_0$, which is preserved as a commutative variable, together with two noncommuting spatial coordinates $\hat{x}_1$ and $\hat{x}_2$. As usual, if the quantum deformation parameter $\xi$ vanishes, the `classical' commutative Galilean spacetime is recovered.

In fact, we stress that~\eqref{spacetime1} is just the noncommutative Galilean spacetime given in Eq.~(5.24) in~\cite{Bernd1}. Thus, we are able to solve one of the open questions posed there: since we have used a kinematical basis for the full DD, this makes it possible to identify the nature of the $R$ generator in Eq.~(5.24) in~\cite{Bernd1}, which turns out to be the quantum  local coordinate associated with the mass generator. Moreover, note that, in full agreement with~\cite{Bernd1}, the noncommutative spacetime algebra~\eqref{spacetime1} is isomorphic to the centrally extended Euclidean Lie algebra, and the noncommutative coordinate $\hat{m}\equiv R$ associated with the mass generator is represented by a compact operator with discrete spectrum.

\subsection{Centrally extended (2+1)-Newton-Hooke Lie algebras as Drinfel'd\\ doubles}
\label{NHDD}

Having shown in the previous section that the two-fold centrally extended (2+1)-Galilei Lie algebra $\overline{\mathfrak{g}(2+1)}$ possesses a DD  structure, we shall generalise this construction to the NH  cases by introducing a non-trivial $\overline{\mathfrak{e}(2)}$ Lie bialgebra structure associated with a quantum deformation. As a result,  the quantum parameter $z$ for the $\overline{\mathfrak{nh}_{\omega}(2+1)}$  deformation will be shown to play the role of the cosmological constant in the outcoming DD Lie algebra in the form $\Lambda=z^2$. This construction reproduces exactly the one in the Lorentzian context described in~\cite{BHMplb}, where the initial trivial Lie bialgebra $\left(\mathfrak{sl}(2),\delta_{0}\right)$ gave rise to the (2+1)-dimensional Poincar\'e algebra as the associated DD Lie algebra, and where the introduction of a quantum deformation of $\mathfrak{sl}(2)$ generated a DD Lie algebra that was shown to be isomorphic to the (2+1) (anti)-de Sitter Lie algebras, and the non-vanishing cosmological constant was directly related to the $\mathfrak{sl}(2)$ quantum deformation parameter $\eta$ through $\Lambda=- \eta^2$.

To this end, let 
us consider the following quantum deformation of $\overline{\mathfrak{e}(2)}$~\cite{Bernd2,BHOS93}: 
\begin{align}
\left[X_1,X_2\right]&=\frac{\displaystyle \sinh (z X_4)}{\displaystyle z}\,, & [X_1,X_3]&=-X_2\,, & [X_2,X_3]&=X_1\,,& [X_4,\,\cdot\,]&=0\,, 
\nonumber 
\end{align}
with coproduct 
\begin{align}
\Delta (X_1)&=X_1\otimes \mathrm{e}^{z X_4}+\mathrm{e}^{-z X_4}\otimes X_1\,, & \Delta (X_2)&=X_2\otimes \mathrm{e}^{z X_4}+\mathrm{e}^{-z X_4}\otimes X_2\,, \nonumber \\
\Delta (X_3)&=X_3\otimes 1+1\otimes X_3\,, & \Delta (X_4)&=X_4\otimes 1+1\otimes X_4\,, \nonumber 
\end{align}
and associated Lie bialgebra structure 
\begin{align}
\delta (X_1)&=z X_1\wedge X_4\,, & \delta (X_2)&=z X_2\wedge X_4\,, & \delta (X_3)&=\delta(X_4)=0\,, & \nonumber
\end{align}
which is generated by the classical $r$-matrix $r=z X_1\wedge X_2$.

The DD Lie algebra~\eqref{agd} associated with this Lie bialgebra is given by
\begin{align}
[X_1,X_2]& =X_4\,, & [X_1,X_3]&=-X_2\,, & [X_2,X_3]&=X_1\,, & [X_4,\cdot]&=0\, , \nonumber
\nonumber \\
\left[x^1,x^4\right]&=z \, x^1\,, & \left[x^2,x^4\right]&=z\, x^2\,, & \left[x^1,x^2\right]&=0\,, & \left[x^3,\cdot\right]&=0\,, \nonumber \\
\left[x^1,X_1\right]&=-z X_4\,, & \left[x^1,X_2\right]&=x^3\,, & \left[x^1,X_3\right]&=-x^2\,, & \left[x^1,X_4\right]&=0\,,  \nonumber \\
\left[x^2,X_1\right]&=-x^3\,, & \left[x^2,X_2\right]&=-z X_4\,, & \left[x^2,X_3\right]&=x^1\,, & \left[x^2,X_4\right]&=0\,,  \nonumber \\
\left[x^3,X_1\right]&=0\,, & \left[x^3,X_2\right]&=0\,, & \left[x^3,X_3\right]&=0\,, & \left[x^3,X_4\right]&=0\,,  \nonumber \\
\left[x^4,X_1\right]&=x^2+z X_1\,, & \left[x^4,X_2\right]&=-x^1+z X_2\,, & \left[x^4,X_3\right]&=0\,, & \left[x^4,X_4\right]&=0\,,  \label{ddhnh}
\end{align}
which is a $z$-deformation of the DD Lie algebra~\eqref{e2e}-\eqref{ddg}.
By setting  
\begin{align}
X_1 & \equiv K_1\,, & X_2 & \equiv K_2\,, & X_3 & \equiv J\,, & X_4 & \equiv S\,, \nonumber \\
x^1 & \equiv -P_2+z K_2\,, & x^2 & \equiv P_1-z K_1\,, & x^3 & \equiv M\,, & x^4 & \equiv -P_0 \nonumber 
\end{align}
one readily checks that the Lie algebra~\eqref{ddhnh} is just $\overline{\mathfrak{nh}_{\omega}(2+1)}$ (\ref{cd}):
\begin{align}
[J,K_i]&=\epsilon_{ij}K_j\,, & [K_1,K_2]&=\sss\,, & [K_i,P_0]&=P_i\,, \nonumber  \\[8pt] 
[J,P_i]&=\epsilon_{ij}P_j\,, & [P_1,P_2]&=-z^2  S\,, & [P_0,P_i]&=-z^2 K_i\,, \nonumber  \\[8pt]
[J,P_0]&=0\,, & [K_i,P_j]&=\delta_{ij}\mm\,, & [\mm,\,\cdot\,]&=[\sss,\,\cdot\,]=0\,, \nonumber
\end{align}
provided that
\be
z ^2= -\omega=\Lambda\,. \label{zz}
\ee

Therefore, according to (\ref{bc}) (see Table~\ref{table1}) we find that

\begin{itemize}

\item In the oscillating NH case with $\omega>0$,   $z$ is  a purely {\em imaginary} deformation  parameter. In terms of the radius $\tau$, this is given by $z={\rm i}/\tau$. 

\item In the expanding NH case with $\omega<0$,   $z$ is  a {\em real}  deformation  parameter such that $z=1/\tau$.

\end{itemize}

 As we will see in section 6,   these relations will be fully consistent with the results obtained by considering the non-relativistic limit of the AdS$_\omega$ algebras studied in~\cite{BHMplb}.

Associated with this double is the canonical classical $r$-matrix (\ref{rmat}), that in this case reads (after multiplying it for the global quantum deformation parameter $\xi$):
\be
\label{rmatrixNHDD}
r_{\xi}=\xi\,r^{\prime}=\xi\big(z K_2\wedge K_1 + \tfrac{1}{2}(K_1\wedge P_2+P_1\wedge K_2+M\wedge J+S\wedge P_0)\big)\,
\ee
which generates the following $\overline{\mathfrak{nh}_{\omega}(2+1)}$ Lie bialgebra cocommutator map:
\begin{align}
\delta _{\xi}(J) & =\delta _{\xi}(S)=\delta _{\xi}(M)=0\,, \nonumber \\
\delta _{\xi}(K_1) & = \xi z S\wedge K_1\,, \nonumber \\
\delta _{\xi}(K_2) & = \xi z S\wedge K_2\,,  \nonumber \\ 
\delta _{\xi}(P_1) & = \xi \left(z M\wedge K_2 +P_2\wedge M -z ^2 K_1\wedge S\right)\,, \label{nhdelta} \\
\delta _{\xi}(P_2) & = \xi \left(z K_1\wedge M -P_1\wedge M -z ^2 K_2\wedge S\right)\,, \nonumber \\
\delta _{\xi}(P_0) & = \xi \left(z K_1\wedge P_2+z P_1\wedge K_2+P_2\wedge P_1 -z ^2 K_1\wedge K_2\right)\,. \nonumber
\end{align}
As it can be immediately checked, this DD structure of the NH  algebras is a cosmologcal constant deformation of the previous Galilei one (\ref{coco1}), which can smoothly be recovered in the $z\to 0$ limit.
 As expected, the $r$-matrix~\eqref{rmatrixNHDD} is (up to a global constant factor) the skew-symmetric counterpart of the one given in Eq.~(4.20) in~\cite{Bernd2}, under the following transformation between the two bases and the cosmological constant parameters:
$$
J\rightarrow - J,
\qquad
J_i\rightarrow\epsilon_{ij}\,K_j,
\qquad
P_i\rightarrow P_i,
\qquad
z=-\sqrt{-\lambda}.
$$

As a consequence, the first-order noncommutative spacetime  coming from~\eqref{nhdelta} is
\begin{align}
\label{spacetime2}
[\hat{x}_1,\hat{x}_2]&=-\xi\hat{x}_0\,, & [\hat{x}_0,\, \cdot \,]&=0\,,
\end{align}
which, as emphasised above, possesses a (commuting) absolute time but, nonetheless, \emph{noncommuting} spatial coordinates. It is worth emphasising the 
fact that Galilei and NH  Lie algebras share the same first-order noncommutative spacetime, which means that the effects of curvature ({\em i.e.} of $\Lambda$)
only enter higher-order terms, as we shall prove in the following section by explicitly constructing the full quantum group. For the sake of completeness, we present now the remaining expressions for the first-order quantum NH$_{\omega}$  group, which are obtained by dualising~\eqref{nhdelta}:
\begin{align}
[\hat{x}_1,\hat{\vartheta}_2]&=\xi z\hat{x}_0=-[\hat{x}_2,\hat{\vartheta}_1]\,, & [\hat{x}_i,\hat{\vartheta}_i]&=0\,, \nonumber \\
[\hat{x}_i,\hat{m}]&=-\xi\epsilon _{ij}\hat{x}_j\,, & [\hat{x}_i,\hat{s}]&=0\,, \nonumber \\
\label{cocoNH} [\hat{\vartheta}_1,\hat{\vartheta}_2]&=-\xi z ^2\hat{x}_0\,, & [\hat{\vartheta}_0,\cdot]&=0\,,  \\
[\hat{\vartheta}_i,\hat{m}]&=\xi z\epsilon _{ij}\hat{x}_j\,, & [\hat{\vartheta}_i,\hat{s}]&=-\xi z\left(\hat{\vartheta}_i+z\hat{x}_i\right)\,, \nonumber \\
[\hat{s},\hat{m}]&=0\, , &  i,j&=1,2.\nonumber
\end{align}
Despite being also first-order relations, the cosmological constant \emph{does} explicitly appear in \eqref{cocoNH}; as expected, these relations lead to the Galilei relations~\eqref{spacetime1} in the vanishing cosmological constant $z\to 0$ limit.


\sect{Quantum groups and noncommutative Galilean spacetimes}

In this Section we face the problem of constructing the full quantum groups corresponding to the (2+1) Galilei and NH  quantum deformations generated by the DD structures that we have just obtained. We are mainly interested in obtaining the noncommutative spacetimes with cosmological constant that will arise when the all-orders quantum deformation is explicitly constructed. To this end, we will follow the same procedure used in~\cite{BHMplb}: firstly, we will construct the full PL group associated with the canonical DD $r$-matrix~\eqref{rmatrixNHDD} that generates the quantum deformation in terms of the local coordinates on the group and, secondly, the quantisation of such Poisson-Hopf algebra will be performed. The noncommutative spacetime will be thus obtained as the subalgebra defined by the spacetime operators that correspond to the classical spacetime local coordinates, whose first order will be given by~\eqref{spacetime2}. As we will see in the sequel, the quantisation of such PL structure can be completed with no ordering ambiguities, and the full quantum group can easily be obtained. This quantum group would be dual, as a Hopf algebra, to the quantum double sketched in~\cite{Bernd2}, but our computations are always performed in a purely kinematical basis.

\subsection{The Poisson-Lie group}

As is well-known, the dual of the cocommutator map $\delta _{\xi}$ of a given Lie bialgebra $(\mathfrak{a},\delta _{\xi})$ provides the linearisation of the PL  group that is uniquely associated with the given Lie bialgebra via the Drinfel'd theorem~\cite{Drib}. Thus, the quantisation of this PL group is the quantum group, $G_{\xi}(\mathfrak{a})$, which is dual, as a Hopf algebra, to the quantum universal enveloping algebra $U_{\xi}(\mathfrak{a})$, the first-order deformation of its corresponding coproduct being given by the cocommutator $\delta _{\xi}$. 

If the Lie bialgebra $(\mathfrak{a},\delta _{\xi})$ is a coboundary one generated by a classical $r$-matrix, 
 the full PL structure is given by the Sklyanin bracket
\be
\{f,g\}=r^{ij}\Big(X_i^L f X_j^L g-X_i^R f X_j^R g\Big)\,,\qquad f,g \in C^{\infty}(G)\,,
\label{sklyan}
\ee
where $X^L\,,X^R$ are the left- and right-invariant vector fields on the group $G$, and the group law on $G$ provides the (co)multiplication under which the bracket~\eqref{sklyan} is invariant.

In our case, $G\equiv\overline{\rm NH_{\omega}(2+1)}$, and its group law was explicitly given in~\cite{Olmo}. From it, left- and right-invariant vector fields on $\overline{\rm NH_{\omega}(2+1)}$ can readily be obtained. Moreover, such group multiplication can be rewritten in Hopf-algebraic terms as a coproduct map $\Delta : C^{\infty}(G)\rightarrow C^{\infty}(G)\otimes C^{\infty}(G)$ in terms of the deformation parameter $z$ (\ref{zz}), which reads
\begin{align}
\Delta (x_0) &=x_0\otimes 1+1\otimes x_0\,,\qquad \qquad \Delta (\vartheta _0) =\vartheta _0\otimes 1+1\otimes \vartheta _0\,, \nonumber \\[2pt]
\Delta (x_1) &=x_1\otimes\Cz (x_0 )+ \vartheta _1\otimes \Sz (x_0 )+\cos (\vartheta _0)\otimes x_1-\sin (\vartheta _0)\otimes x_2\,, \nonumber \\[2pt]
\Delta (x_2) &=x_2\otimes\Cz (x_0 )+ \vartheta _2\otimes \Sz (x_0 )+\sin (\vartheta _0)\otimes x_1+\cos (\vartheta _0)\otimes x_2\,, \nonumber \\[2pt]
\Delta (\vartheta _1) &=\vartheta _1\otimes\Cz (x_0 )+ z^2 x _1\otimes \Sz (x_0 )+\cos (\vartheta _0)\otimes \vartheta _1-\sin (\vartheta _0)\otimes\vartheta _2\,, \nonumber \\[2pt]
\Delta (\vartheta _2) &=\vartheta _1\otimes\Cz (x_0 ) +z^2 x _2\otimes \Sz (x_0 )+\sin (\vartheta _0)\otimes\vartheta _1+\cos (\vartheta _0)\otimes\vartheta _2\,, \label{coproduct}  \\
 \Delta (s) &=s\otimes 1+1\otimes s -\frac{z^2}{2}\left\{\Big(\cos\vartheta _0\big(x_1\otimes x_2-x_2\otimes x_1-\frac{1}{z^2}(\vartheta _1\otimes\vartheta _2-\vartheta _2\otimes\vartheta _1)\big)\Big.\right. \nonumber \\
                  &+\left.\Big.\sin\vartheta _0\big(x_1\otimes x_1+x_2\otimes x_2-\frac{1}{z^2}\left(\vartheta _1\otimes\vartheta _1+\vartheta _2\otimes\vartheta _2\right)\big) \Big)      \Cz (x_0 )\right. \nonumber \\
                  &+  \Big(\cos\vartheta _0\big(x_2\otimes \vartheta _1-x_1\otimes\vartheta _2+\vartheta _1\otimes x _2-\vartheta _2\otimes x _1\big)\Big. \nonumber \\
                  &\left.\Big.-\sin\vartheta _0\big(x_1\otimes\vartheta _1+x_2\otimes\vartheta _2-\vartheta _1\otimes x _1-\vartheta _2\otimes x _2\big)\Big)  \Sz (x_0 )\right\}\,, \nonumber \\
\Delta (m)&=m\otimes 1+1\otimes m+ z^2 \left(x_1\vartheta _1\otimes 1+x_2\vartheta _2\otimes 1\right) \Sz^2 (x_0 )\nonumber \\
& +\frac{1}{4}\Big(  \vartheta _1^2\otimes 1+\vartheta _2^2\otimes 1 + z^2\big(x_1^2\otimes1+x_2^2\otimes 1\big)\Big)\  \Sz (2x_0 )\nonumber \\
                 &    +\Big(\vartheta _1\big(\cos (\vartheta _0)\otimes x_1-\sin (\vartheta _0)\otimes x_2\big) +\vartheta _2\big(\sin (\vartheta _0)\otimes x_1+\cos (\vartheta _0)\otimes x_2\big)\Big)  \Cz (x_0 ) \nonumber \\
                 &+z^2 \Big(x_1\big(\cos (\vartheta _0)\otimes x_1-\sin (\vartheta _0)\otimes x_2\big) 
                +x_1\big(\cos (\vartheta _0)\otimes x_1-\sin (\vartheta _0)\otimes x_2\big)\Big) \Sz (x_0 )\,, \nonumber
\end{align}
where, for the sake of convenience, we have defined the functions
$$
\Cz(x_0)\equiv \cosh(z\,x_0)\,, \qquad \Sz(x_0)\equiv \frac{\sinh(z\, x_0)}{z}\,.
$$
This way, the explicit  coproduct  for  the oscillating NH   ($z={\rm i}/\tau$) and expanding NH  ($z={1}/\tau$)  cases, along with their vanishing cosmological Galilean   limits  ($z=0;\, \tau\to \infty$), can be straightforwardly obtained by taking into account that
\begin{equation}
\Cz (x_0)  =\left\{
\begin{array}{ll}
 \cos\left(\frac{x_0}{\tau}\right)&\   z= {\rm i}/\tau \\[2pt] 
  1  &\ 
  z=0 \\[2pt]
 \cosh\left(\frac{x_0}{\tau}\right)&\   z= {1}/\tau 
\end{array}\right.\  ,\qquad
\Sz (x_0)  =\left\{
\begin{array}{ll}
\tau \sin\left(\frac{x_0}{\tau}\right)&\   z= {\rm i}/\tau \\[2pt] 
  x_0  &\ 
  z=0 \\[2pt]
 \tau \sinh\left(\frac{x_0}{\tau}\right)&\   z= {1}/\tau 
\end{array}\right.   .
\nonumber
\end{equation}

Thus, the full Poisson-Hopf algebra on $C^\infty\left(\overline{\rm NH_{\omega}(2+1)}\right)$, with cosmological constant $z ^2=\Lambda$ (\ref{zz}), 
is given by the coproduct~\eqref{coproduct} and the Sklyanin bracket~\eqref{sklyan} among the eight local group coordinates $\{x_a,m,\vartheta _a,s\}$ $(a=0\,,1\,,2)$ associated with the classical $r$-matrix~\eqref{rmatrixNHDD}.

The Poisson subalgebra generated by the spacetime coordinates is given by
\be
\{x_1,x_2\} =-\xi\Sz (x_0 )\Upsilon _{z}(x_0)\,, 
\qquad
\{x_0,x_1\}=\{x_0,x_2\}=\,0\,,
\label{pst}
\ee
where
\be
\Upsilon _{z}(x_0)\equiv\Cz   ({x_0} )-z\Sz ({x_0} )\, .
\ee
Note that the time coordinate $x_0$ Poisson-commutes with the spatial ones, and a series expansion of this bracket in terms of the cosmological constant parameter $z$ reads
$$
\label{series1} \{x_1,x_2\}=-\xi x_0+\xi z x_0^2-\tfrac{2}{3}\xi z ^2x_0^3+\tfrac{1}{3}\xi z ^3 x_0^4+\mathcal{O}(x_0^5)\,,
$$
whose first order reproduces the Poisson version of \eqref{spacetime2}, as it should be. On the other hand,    the vanishing cosmological constant limit $z= 0$ ($\tau\to\infty$) leading  to the (2+1) Galilean PL  structure   has no higher-order corrections.

Therefore,~\eqref{pst} provides the full non-vanishing cosmological constant expression for the Poisson spacetime associated with the DD structure of the $\overline{\mathfrak{nh}_{\omega}(2+1)}$ algebra here presented. According to the value of the deformation parameter $z$,  we obtain the following  explicit PL brackets:
\begin{equation}
\begin{array}{lll}
 \displaystyle{ \overline{\mathfrak{nh}_{+}(2+1)}\   \ (z={\rm i}/\tau)  }:&\    \{x_1,x_2\} =-\xi\tau \sin\left(\frac{x_0}{\tau}\right)
 \left\{   \cos\left(\frac{x_0}{\tau}\right)   -{\rm i}   \sin\left(\frac{x_0}{\tau}\right)\right\},
  &\ 
   \{x_0,x_i\}= 0\, .  \\[8pt] 
    \displaystyle{ \overline{\mathfrak{g}(2+1)}\   \ (z=0)  }:&\    \{x_1,x_2\} =-\xi \,    {x_0} ,
  &\ 
   \{x_0,x_i\}= 0\, .  \\[8pt] 
  \displaystyle{ \overline{\mathfrak{nh}_{-}(2+1)}\   \ (z=1/\tau)  }:&\    \{x_1,x_2\} =-\xi\tau \sinh\left(\frac{x_0}{\tau}\right)
 \left\{   \cosh\left(\frac{x_0}{\tau}\right)   -   \sinh\left(\frac{x_0}{\tau}\right)\right\},
  &\ 
   \{x_0,x_i\}= 0\, .  \\[8pt] 
 \end{array}
\label{cdddd}
\end{equation}

The remaining PL  brackets coming from the Sklyanin bracket~\eqref{sklyan} read $(i,j=1,2)$
\begin{align}
\{x_0,\,\cdot \,\}&=\,0\,,  \nonumber \\
\{x_1,\vartheta _2\} &=\xi z\Sz (x_0 )\Upsilon _{z}(x_0)=-\{x_2,\vartheta _1\}\,, \nonumber \\
 \{x_i,\vartheta _i\}&=\,0\,,\nonumber \\
\{x_i,m\} &=-\xi\epsilon _{ij}x_j\Cz ({x_0} )\Upsilon _{z}(x_0)\,, \nonumber \\
\label{PL} \{x_i,s\} &=-\tfrac{1}{2}\xi z (z x_i+\vartheta _i)\Sz (x_0 )\Upsilon _{z}(x_0)\,, \\
\{\vartheta _0,\, \cdot \,\}&=\,0\,, \nonumber \\
\{\vartheta _1,\vartheta _2\} &=-\xi z ^2\Sz (x_0 )\Upsilon _{z}(x_0)\,, \nonumber \\
\{\vartheta _i,m\} &=\xi z\, \epsilon _{ij}x_j\Cz ({x_0} )\Upsilon _{z}(x_0)  \,, \nonumber \\
\{\vartheta _i,s\} &=-\tfrac{1}{2}\xi z (z x_i+\vartheta _i)\left(1+\Cz ({x_0} )\Upsilon _{z}(x_0)\right)\,, \nonumber \\
\{s,m\} &=\tfrac{1}{2}\xi z\Cz ({x_0} )\Big(z \left(x_1^2+x_2^2\right) + x_1\vartheta _1+x_2\vartheta _2\Big)\Upsilon _{z}(x_0)\,.\nonumber 
\end{align}
It is readily checked that the linearisation of~\eqref{PL} in terms of the local coordinates gives as a result the Poisson version of (\ref{cocoNH}). It is worth pointing out that $\{x_0,x_1,x_2\}$ generate a Poisson subalgebra which is {\em not} a Hopf subalgebra (see the coproducts~\eqref{coproduct} for these local coordinates).

\subsection{Quantum $\overline{\rm NH_{\omega,\xi}(2+1)}$ group and noncommutative spacetime}

The quantisation, as a Hopf algebra,  of the PL group  described above is just the quantum $\overline{\rm NH_{\omega,\xi}(2+1)}$ group  dual to $U_{\xi}\Big(\overline{\mathfrak{nh}_{\omega}(2+1)}\Big)$ we are seeking. This quantisation turns out to be immediate, since $x_0$ Poisson-commutes with all the remaining local coordinates and $\{x_i,\vartheta _i\}=0$. This implies that neither the PL brackets~\eqref{pst} and~\eqref{PL} nor the coproduct map~\eqref{coproduct} have any kind of ordering ambiguities.

Therefore, the quantisation can be performed by promoting the eight commutative group coordinates to the quantum, noncommutative operators dual to the quantum algebra generators (see~\eqref{pairing}) and by replacing the Poisson brackets by commutation rules. 
Afterwards,  the algebra homomorphism condition between the coproduct and the commutation rules can easily be checked.

In particular, in order to highlight the associated quantum spacetime, let us explicitly write down the quantum analogue of~\eqref{pst},
\begin{align}
\left[\hat{x}_1,\hat{x}_2\right]&=-\xi\Sz ({\hat{x}_0} )\Upsilon _{z}(\hat{x}_0)
=-\xi \big(\hat{x}_0 - z \hat{x}_0^2+\tfrac{2}{3}z ^2\hat{x}_0^3+\mathcal{O}\left(\hat{x}_0^4\right)\big)\,, & \left[\hat{x}_0,\, \cdot\, \right]=\,0\,,
\label{ncst}
\end{align}
which would be the non-relativistic noncommutative  spacetime with cosmological constant associated with the DD structure of the corresponding centrally extended kinematical groups. Indeed, the fact that $\hat{x}_0$ is a central operator in the full quantum group implies that it can be considered as a multiple of the identity on each irreducible representation of the quantum double. 

 It is worth stressing that the commutator version of the brackets~\eqref{pst} and~\eqref{PL}, together with the coproduct map~\eqref{coproduct}, provides the full quantum $\overline{\rm NH_{\omega,\xi}(2+1)}$ group, whose construction was partially given in~\cite{Bernd2} by making use of a different basis for both the Lie algebra generators and the corresponding group coordinates. This explains why the noncommutative spacetime~\eqref{ncst} above and the one given in Eq.~(6.4) in~\cite{Bernd2} are given by different power series in the time generator $\hat{x}_0$. Nevertheless, in both cases, the leading term in both series is the same (a constant time  $\hat{x}_0$), as it should be.


\sect{The non-relativistic limit of (2+1) Lorentzian doubles}

In this section, our aim is to present the non-relativistic limit of all the AdS$_{\omega}(2+1)$ canonical classical $r$-matrices arising from all their possible DD structures that were explicitly constructed in~\cite{BHMcqg}. 
As a result, we shall obtain all of $\overline{\mathfrak{nh}_{\omega}(2+1)}$ classical $r$-matrices that can be obtained as contractions of the (trivially extended)  AdS$_{\omega}(2+1)$ ones, and we will explicitly show that \eqref{rmatrixNHDD} is just the non-relativistic limit of the Lorentzian DD $r$-matrix that has thoroughly been studied in~\cite{BHMplb}. For the sake of self-consistency, we will also show that the non-relativistic limit of the noncommutative spacetime associated with such Lorentzian $r$-matrix presented in~\cite{BHMplb} can also be performed, and this group contraction procedure just leads to the Galilean Poisson spacetime~\eqref{cdddd} associated with the $r$-matrix \eqref{rmatrixNHDD}.


\subsect{Non-relativistic limit of  Lorentzian $r$-matrices}

We recall that, as was shown in~\cite{Witten1}, the    three Lorentzian Lie algebras of  the isometry groups  of  Lorentzian (2+1)-gravity can be comprised within AdS$_\omega$, a one-parameter structure that includes
  $\mathfrak{so}(2,2)$ $(\k>0)$, $\mathfrak{iso}(2,1)$ $(\k=0)$ and  $\mathfrak{so}(3,1)$ $(\k<0)$,  in which  the cosmological constant $\Lambda=-\k$ plays the role of a structure constant. In terms of the kinematical  generators  
of translations  $T_a$  $(a=0,1,2)$  and  Lorentz transformations $J_a$  $(a=0,1,2)$, the Lie brackets  for  AdS$_\omega$ take the form 
\be 
\begin{array}{lll}
 [J_0,J_1]=J_2,  &\qquad [J_0,J_2]=-J_1,  &\qquad [J_1, J_2]=-\,J_0, \\[2pt]
 [J_0,T_0]=0 ,   &\qquad [J_0,T_1]= T_2 ,  &\qquad [J_0, T_2]= - T_1,\\[2pt]
[J_1,T_0]=-T_2 ,  &\qquad [J_1,T_1]=0 ,  &\qquad [J_1, T_2]= -\,T_0,\\[2pt]
[J_2,T_0]=T_1 ,   &\qquad [J_2,T_1]= \,T_0 ,   &\qquad [J_2, T_2]=0,\\[2pt]
[T_0,T_1]=-\Lambda    J_2,  &\qquad [T_0, T_2]= \Lambda  J_1  , &\qquad [T_1,T_2]= \Lambda    J_0   .
 \end{array}
  \label{da}
\ee 

Recently,  a classification  of all possible $D(\mathfrak{g})$ structures for the  (2+1)-dimensional AdS and dS algebras which are compatible with the requirements of (2+1)-gravity has been performed in~\cite{BHMcqg}. This classification leads  to {\em seven} possible canonical classical $r$-matrices (four for dS and three for AdS) that generate the corresponding quantum deformations, which can be considered as suitable candidates for symmetries of (2+1) quantum gravity models. However, such classification is reduced to  only {\em four} classical $r$-matrices if we further impose the existence of a well-defined Poincar\'e/Minkowskian flat limit $(\Lambda\to 0)$, along with the existence of the appropriate pairing for a   CS formulation of (2+1)-gravity~\cite{MSquat, MS}, that is,
\be
 \langle J_a,T_b\rangle =g_{ab}, \qquad \langle J_a,J_b\rangle =0, \qquad \langle T_a,T_b\rangle =0    ,\qquad a,b=0,1,2,
   \label{db}
   \ee
with metric $g=\rm{diag}(-1,1,1)$.

These are the ${\bf dS}^{2+1}$ $r$-matrices A and C, together with the ${\bf AdS}^{2+1}$ ones, E and F, displayed in Table~\ref{table2},  where it is important to recall that $\eta$ is a {\em real} parameter related to the cosmological constant (that plays a similar role to 
 $z$):
\begin{align}
{\bf AdS}^{2+1}&: & \eta &=\pm\frac{1}{\tau}\,, & \k&=\eta^2=\frac 1{\tau^2}>0\,, & \Lambda&=- \k <0\,; \nonumber\\
{\bf dS}^{2+1} &:   & \eta &=\pm \frac{{1}}{\tau}\,, &  \k&=-\eta^2=-\frac 1{\tau^2}<0\,,  &   \Lambda&=- \k  >0\,. \label{dc}
\end{align}
In particular, note that both signs for $\eta$ are allowed. At this point it is important to stress (see~\cite{BHMcqg,Naranjo} for an exhaustive discussion) that the $r$-matrices C and F are twisted versions of the $\kappa$-deformation of the AdS$_\omega$ algebra, whereas A and E define a completely different quantum deformation, that was presented for the first time in~\cite{BHMplb10}.

Now, the central extension of the AdS$_\omega$ algebra~\eqref{da} can be obtained by introducing two new central generators $M$ and $S$ and by performing the change of basis
\begin{align}
P_0 &=T_0-c^2 \mm\,, & P_1 &=\frac 1c\, T_1\,, & P_2 &=\frac 1c\, T_2\,,  \nonumber \\[2pt]
K_1 &=\frac 1c\, J_2\,, & K_2 &=-\frac 1c\, J_1\,, & J &=J_0+c^2 S\,, \label{dd}      
\end{align}
such that~\eqref{da} are transformed into \eqref{cb}. Afterwards, if we introduce the change (\ref{dd}) in the four Lorentzian  $r$-matrices of Table~\ref{table2} and apply the non-relativistic limits defined by
\bea
&& r_{\rm A1}=\lim_{c\to \infty} \frac1{c^2}\left( r'_{\rm A}-\frac 12 c^4 M\wedge S  \right) ,\qquad  r_{\rm C1}=\lim_{c\to \infty} \frac1{c^3} r'_{\rm C} ,\label{ra1}\\[4pt]
&& r_{\rm E1}=\lim_{c\to \infty} \frac1{c^3}\left( r'_{\rm E}-\frac 12 c^4 M\wedge S  \right) ,\qquad  r_{\rm F1}=\lim_{c\to \infty} \frac1{c^3} r'_{\rm F} ,\label{re1}
\eea
 we obtain the  classical $r$-matrices A1, C1, E1 and F1 for the extended Galilean algebras $\overline{\mathfrak{nh}_\k(2+1)}$ with commutation relations (\ref{cd}), which are also displayed in Table~\ref{table2}, and where the parameter  $\eta$ is defined in a similar way to (\ref{dc}):
\begin{align}
{\bf NH}_+^{2+1}& : & \eta &=\pm\frac{1}{\tau}\,, &  \k&=\eta^2=\frac 1{\tau^2}>0\,, &  \Lambda &=- \k <0\,  ; \nonumber\\
{\bf NH}_-^{2+1} &: &  \eta &=\pm \frac{{1}}{\tau}\,, & \k&=-\eta^2=-\frac 1{\tau^2}<0\,, &  \Lambda &=- \k   >0\,. \label{de}
\end{align} 
Note that the term in~\eqref{ra1} and~\eqref{re1} containing $M\wedge S$ has to be added to the initial AdS$_\omega$ $r$-matrices in order to guarantee the convergence of the limiting procedure. This can be done because both $M$ and $S$ are central generators and, therefore, the term $M\wedge S$ is trivial as an $r$-matrix, since it does not contribute to the corresponding cocommutator~\eqref{rcanon2}. 

Moreover, there exists a second possibility (in this case, involving complex coefficients) that enables us to relate the commutation relations (\ref{da}) and (\ref{cb}) and, at the same time, fulfilling (\ref{dc}), namely
 \begin{align}
P_0&={\rm i}T_1-c^2 M\,, & P_1&=-\frac 1c\, T_2\,, & P_2&=\frac {\rm i}c\, T_0\,, \nonumber \\[2pt]
\label{df} J&= {\rm i}J_1+c^2 S\,, & K_1&=\frac  {\rm i}c\, J_0\,, & K_2&=\frac 1c\, J_2\,.             
\end{align}
Now, substituting (\ref{df}) in the four Lorentzian  $r$-matrices of Table~\ref{table2} and applying the non-relativistic limits given by
\bea
&& r_{\rm A2}=\lim_{c\to \infty} \frac1{c^3}\left( r'_{\rm A}-\frac 12 c^4 M\wedge S  \right) ,\qquad  r_{\rm C2}=\lim_{c\to \infty} \frac1{c^3} r'_{\rm C} ,\nonumber\\[4pt]
&& r_{\rm E2}=\lim_{c\to \infty} \frac1{c^2}\left( r'_{\rm E}-\frac 12 c^4 M\wedge S  \right) ,\qquad  r_{\rm F2}=\lim_{c\to \infty} \frac1{c^3} r'_{\rm F} ,\nonumber
\eea
yield a second set (A2, C2, E2 and F2) of   classical $r$-matrices for $\overline{\mathfrak{nh}_{\omega}(2+1)}$, which are also explicitly given in Table~\ref{table2}, and where the parameter  $\eta$ is also given by (\ref{de}).

Now, by recalling that the non-relativistic DD classical $r$-matrix~\eqref{rmatrixNHDD} that we have presented in section 4 (and firstly obtained in \cite{Bernd1,Bernd2} in a different basis) is given by
$$
r^{\prime}=z K_2\wedge K_1 + \tfrac{1}{2}(K_1\wedge P_2+P_1\wedge K_2+M\wedge J+S\wedge P_0)\equiv r_{\rm NH}^{\rm DD}\,, 
$$
the information contained in Table~\ref{table2} makes it immediate to identify this classical $r$-matrix as the non-relativistic limit of a Lorentzian $r$-matrix.
Indeed, since the cases A1 and E2 read 
\bea
&& {\bf NH}_-^{2+1} :\quad  r_{\rm A1}=\eta K_1\wedge K_2+\tfrac 1 2 (K_2\wedge P_1-K_1\wedge P_2+P_0\wedge S+J\wedge M) ,\nonumber\\[2pt]
&& {\bf NH}_+^{2+1} :\quad  r_{\rm E2}=-{\rm i}\eta K_1\wedge K_2+\tfrac 1 2 (K_2\wedge P_1-K_1\wedge P_2+P_0\wedge S+J\wedge M),
\label{re2}
\eea
we get that for the expanding NH  case, ${\bf NH}_-^{2+1}$ $(z=1/\tau)$,
\be
r'=-r_{\rm A1}\,, \qquad \mbox{with}\quad \eta= z\equiv 1/\tau\,,
\label{raa1}
\ee
which means that we can consider this $r$-matrix as a contraction of a dS DD structure, whereas for the oscillating NH case, ${\bf NH}_+^{2+1}$ $(z={\rm i}/\tau)$, the $r$-matrix $r'$ arises as a contraction from an AdS DD $r$-matrix in the form
\be
r'=-r_{\rm E2}\,,\qquad \mbox{with}\quad \eta= {\rm i}z \equiv -{1}/\tau\,.
\label{ree2}
\ee

Therefore, we have shown that the DD structure suitable as the symmetry for a CS approach to Galilean (2+1)-gravity with cosmological constant can be understood as the non-relativistic limit of the DD $r$-matrix of the AdS$_\omega$ algebra that has previously been studied in~\cite{BHMplb,BHMplb10}.


\subsect{Non-relativistic limit of  (2+1) noncommutative AdS  spacetimes}

We recall that the Poisson version of the noncommutative AdS$_\omega$ spacetime corresponding to the AdS case E was explicitly constructed in~\cite{BHMplb}, 
 and reads
\bea
&&  \{t_0,t_1\} = -
\xxi\,\frac{\tanh\m t_2 }{\m} 
\,\C,
\qquad
 \{t_0,t_2\} = \xxi\,
 \frac{ \tanh\m t_1}{\m}\,
\C ,
\qquad
 \{t_1,t_2\} = \xxi\,
\frac{\tan\m t_0}{\m}
\,\C ,\nonumber\\[2pt]
&& \C(t_0,t_1)=\cos\m t_0\left (\cos\m t_0\cosh\m t_1+ \sinh\m t_1\right ),
\label{qa}
\eea
where $t_a$ are the group coordinates of the generators $T_a$ ($a=0,1,2$) and $\xxi$ is the (quantum) deformation parameter that is introduced as a multiplicative factor in the classical $r$-matrix. 
We recall that this Poisson structure is not symplectic and  its symplectic leaves are the level surfaces of the function
\be
C(t_0,t_1,t_2) = \cos(\eta t_0) \cosh(\eta t_1) \cosh(\eta t_2),\qquad \{ C, t_a\}=0.
\label{qaa}
\ee
 Since  the classical $r$-matrix E2 (\ref{re2}) of ${\bf NH}_+^{2+1} $ has been obtained  from the AdS one of type E through a non-relativisitc limit, it is expected that an appropriate contraction procedure on the group manifold should give rise to the corresponding non-relativistic noncommutative spacetime.

Explicitly, by taking into account (\ref{df}), (\ref{ree2}) and a necessary scaling of $\xxi$,  we consider the map defined by
$$
x_0=-{\rm i}\,t_1,\qquad x_1=-c\, t_2,\qquad x_2=-{\rm i}c\, t_0,\qquad \eta=-1/\tau,\qquad \xxi=-\xi/c^2 \, .
$$
 Next,  by introducing it into (\ref{qa})  and (\ref{qaa}) we obtain
 \bea
&& \{x_0,x_1\} = \frac{\xi}{c^2}\,\frac{\tanh\left( \frac{x_2}{c\tau}\right)}{1/c\tau}
\,\C, \quad
   \{x_0,x_2\} =- \frac{\xi}{c^2}\,\frac{\tanh\left(   \frac{x_1}{c\tau}   \right)}{1/c\tau}
\,\C, \quad
  \{x_1,x_2\} =-  {\xi} \,\frac{\tan\left( \frac{x_0}{\tau}\right)}{1/\tau}
\,\C,
\nonumber\\[2pt]
&& \C(x_0,x_2)= \cosh\left( \frac{x_2}{c\tau} \right) \left\{  \cosh\left(\frac{x_2}{c\tau} \right) \cos\left( \frac{x_0}{\tau}\right)  -{\rm i}  \sin\left( \frac{x_0}{\tau}\right) \right\},
\nonumber\\[2pt]
&& C(x_0,x_1,x_2)=\cos\left( \frac{x_0}{\tau} \right)\cosh\left( \frac{x_1}{c\tau} \right) \cosh\left( \frac{x_2}{c\tau} \right) .
\nonumber
\eea
The non-relativistic $c\to \infty$ limit gives
 \bea
&& \{x_0,x_1\} = 0, \qquad
   \{x_0,x_2\} =0, \qquad
  \{x_1,x_2\} =-  {\xi} \,\frac{\tan\left( \frac{x_0}{\tau}\right)}{1/\tau}
\,\C,
\nonumber\\[2pt]
 \nonumber\\[2pt]
&& \C(x_0)=   \cos\left( \frac{x_0}{\tau}\right)  -{\rm i}  \sin\left( \frac{x_0}{\tau}\right) ,\qquad  C(x_0)=\cos\left( \frac{x_0}{\tau} \right),
\label{qabc}
\eea
and if we consider the map
\be
x_i'=\sqrt{\cos\left( \frac{x_0}{\tau}\right)}\,x_i, \quad i=1,2,
\qquad
x_0'=x_0,
\ee
 the Poisson algebra generated by $\{x_0',x_1',x_2'\}$ is just the noncommutative Galilean Poisson spacetime~\eqref{cdddd} associated with  $ \overline{\mathfrak{nh}_{+}(2+1)}$,
that has thus been obtained as the appropriate non-relativistic counterpart of~\eqref{qa}. Note that the essential effect of the $c\to\infty$ limit is to provide the emergence of an `absolute' time that commutes with the (albeit noncommutative) Galilean space coordinates $x_1'$ and $x_2'$.

It is also worth stressing that this result is completely different to the one that is obtained if {\em any} of the Galilean $r$-matrices C1, C2, F1 and F2 that arise as the non-relativistic limits of the twisted $\kappa$-AdS$_\omega$ deformations C and F are considered. It is immediate to check that in all these four cases the associated Galilean PL  groups have {\em vanishing} Poisson brackets among the {\em three spacetime coordinates} $(x_0,x_1,x_2)$ (although the full PL  structure involving the remaining local coordinates on the group is nevertheless a non-trivial one). This means that the $c\to \infty$ limit of the noncommutative twisted $\kappa$-AdS$_\omega$ spacetime studied in \cite{Naranjo} gives rise to a `classical' spacetime. Such a strong physical difference between the two possible AdS$_\omega$ quantum deformations that are compatible with the CS approach and coming from DD structures should be relevant in order to elucidate their possible role as symmetries in (2+1) quantum gravity.


\begin{table}[t] {\footnotesize
 \noindent
\caption{{\small The four    Lorentzian DD   $D(\mathfrak{g})$ with quantum  deformation parameter $\eta$ and commutation relations (\ref{da}), providing   a well-defined flat/Poincar\'e   limit $\eta\to 0$ ($\Lambda=0$) as well as  the pairing given by (\ref{db}). Their corresponding non-relativistic centrally extended Galilean  $r$-matrices   with commutation relations (\ref{cd}) are also shown.
  }}
\label{table2}
\medskip
\noindent\hfill
\begin{tabular} {lllllll}
 \hline
 & & & & \\[-1.5ex]
 \#   & $\Lambda$ &   Skew-symmetric $r$-matrix & $D(\mathfrak g)$& Space\\[+1.5ex]
 \hline
 & & & & \\[-1.5ex]
 A &  $\eta^2$ & $r'_{\rm A}=\eta J_1\wedge J_2+\tfrac 1 2 (T_1\wedge J_1+T_2\wedge J_2-T_0\wedge J_0)$ & $so(3,1)$&${\bf dS}^{2+1}$\\[+1.0ex]
      & 0 & $ r'_{\rm A}= \tfrac 1 2 (T_1\wedge J_1+T_2\wedge J_2-T_0\wedge J_0)$ & $iso(2,1)$&${\bf M}^{2+1}$\\[+2.5ex]
 C  & $\eta^2$  &  $r'_{\rm C}=\tfrac 1 2 (J_1\wedge T_0-  J_0 \wedge T_1+J_2\wedge T_2)$   & $so(3,1)$ &${\bf dS}^{2+1}$\\[+1.0ex]
   &     $0$  &  $r'_{\rm C}=\tfrac 1 2 (J_1\wedge T_0-  J_0 \wedge T_1+J_2\wedge T_2)$   & $iso(2,1)$ &${\bf M}^{2+1}$\\[+2.5ex]
   E &  $-\eta^2$  & $r'_{\rm E}=\eta J_0\wedge J_2+\tfrac 1 2 (T_1\wedge J_1+T_2\wedge J_2-T_0\wedge J_0)$ & $so(2,2)$&${\bf AdS}^{2+1}$  \\[+1.0ex]
 &   $0$ & $r'_{\rm E} = \tfrac 1 2 (T_1\wedge J_1+T_2\wedge J_2-T_0\wedge J_0)$ & $iso(2,1)$&${\bf M}^{2+1}$  \\[+2.5ex]
 F   &  $-\eta^2$  & $r'_{\rm F}  =\tfrac 1 2 (J_1\wedge T_0- J_0 \wedge T_1 +J_2\wedge T_2) $   & $so(2,2)$&${\bf AdS}^{2+1}$ \\[+1.0ex]
    &    $0$    & $r'_{\rm F}  =\tfrac 1 2 (J_1\wedge T_0- J_0 \wedge T_1 +J_2\wedge T_2) $   & $iso(2,1)$&${\bf M}^{2+1}$ \\[+2.5ex]
  \hline

   & & & & \\[-1.5ex]
 A1 &  $\eta^2$ & $r_{\rm A1}=\eta K_1\wedge K_2+\tfrac 1 2 (K_2\wedge P_1-K_1\wedge P_2+P_0\wedge S+J\wedge M)$ & $\overline{\cal N}_-$&${\bf NH}_-^{2+1}$\\[+1.0ex]
      & 0 & $ r_{\rm A1}= \tfrac 1 2  (K_2\wedge P_1-K_1\wedge P_2+P_0\wedge S+J\wedge M )$ & $\overline{iiso}(2)$&${\bf G}^{2+1}$\\[+2.5ex]
 C1  & $\eta^2$  &  $r_{\rm C1}=\tfrac 1 2 (S\wedge P_1 +M \wedge K_2)$   & $\overline{\cal N}_-$ &${\bf NH}_-^{2+1}$\\[+1.0ex]
   &     $0$  &  $r_{\rm C1}=\tfrac 1 2 (S\wedge P_1 +M \wedge K_2)$   & $\overline{iiso}(2)$ &${\bf G}^{2+1}$\\[+2.5ex]
   E1 &  $-\eta^2$  & $r_{\rm E1}= -\eta S\wedge K_1$ & $\overline{\cal N}_+$&${\bf NH}_+^{2+1}$  \\[+1.0ex]
 &   $0$ & $r_{\rm E1} = 0 $ & $\overline{iiso}(2)$&${\bf G}^{2+1}$  \\[+2.5ex]
 F1   &  $-\eta^2$  & $r_{\rm F1}  =\tfrac 1 2 (S\wedge P_1 +M \wedge K_2) $   & $\overline{\cal N}_+$&${\bf NH}_+^{2+1}$ \\[+1.0ex]
    &    $0$    & $r_{\rm F1}  =\tfrac 1 2 (S\wedge P_1 +M \wedge K_2) $   & $\overline{iiso}(2)$&${\bf G}^{2+1}$ \\[+2.5ex]
  \hline

   & & & & \\[-1.5ex]
 A2 &  $\eta^2$ & $r_{\rm A2}={\rm i} \eta S\wedge K_2 $ & $\overline{\cal N}_-$&${\bf NH}_-^{2+1}$\\[+1.0ex]
      & 0 & $ r_{\rm A2}=0$ & $\overline{iiso}(2)$&${\bf G}^{2+1}$\\[+2.5ex]
 C2  & $\eta^2$  &  $r_{\rm C2}=\tfrac 1 2 (S\wedge P_2-M \wedge K_1)$   & $\overline{\cal N}_-$ &${\bf NH}_-^{2+1}$\\[+1.0ex]
   &     $0$  &  $r_{\rm C2}=\tfrac 1 2 (S\wedge P_2-M \wedge K_1)$   & $\overline{iiso}(2)$ &${\bf G}^{2+1}$\\[+2.5ex]
   E2 &  $-\eta^2$  & $r_{\rm E2}=-{\rm i}\eta K_1\wedge K_2+\tfrac 1 2 (K_2\wedge P_1-K_1\wedge P_2+P_0\wedge S+J\wedge M)$ & $\overline{\cal N}_+$&${\bf NH}_+^{2+1}$  \\[+1.0ex]
 &   $0$ & $r_{\rm E2} =\tfrac 1 2 (K_2\wedge P_1-K_1\wedge P_2+P_0\wedge S+J\wedge M)$ & $\overline{iiso}(2)$&${\bf G}^{2+1}$  \\[+2.5ex]
 F2   &  $-\eta^2$  & $r_{\rm F2}  =\tfrac 1 2 (S\wedge P_2-M \wedge K_1) $   & $\overline{\cal N}_+$&${\bf NH}_+^{2+1}$ \\[+1.0ex]
    &    $0$    & $r_{\rm F2}  =\tfrac 1 2 (S\wedge P_2-M \wedge K_1) $   & $\overline{iiso}(2)$&${\bf G}^{2+1}$ \\[+2.5ex]
  \hline

 \end{tabular}
\hfill}
\end{table}

\sect{Concluding remarks}

Drinfel'd doubles seem to provide a sound basis to face the quantisation of gravity by linking (2+1) CS gravity and admissible PL structures on the kinematical groups of the underlying model spacetimes. In this article,
we have explicitly constructed the full quantum double associated with (2+1)-Galilean gravity, and analysed the associated quantum spacetime,
elaborating on previous work performed
in~\cite{BHMcqg,Bernd1,Bernd2}. Also, in so doing, we have demonstrated that the
quantum double symmetry for (2+1)-Galilean gravity can be recovered as a well-defined non-relativistic limit from a certain quantum double of (2+1)-Lorenztian gravity, thereby
providing a unified setting to accommodate, on one hand, both relativistic and non-relativistic scenarios and, on the other, both flat and
non-flat cases (in this respect, recall that throughout the whole paper the cosmological constant enters as a deformation parameter). For the sake of completeness, all
classical $r$-matrices originating as non-relativistic contractions of all possible DD structures for (2+1)-Lorenztian gravity given in \cite{BHMcqg} have also been presented.

An interesting fact that emerges in the quantisation framework here analysed is the nature of the Galilean quantum spacetime. It has been emphasised in the body of this paper the
`absoluteness' of the quantum time coordinate, as it turns out to be a central element of the algebra of noncommutative spacetime operators in the non-relativistic scenario. Unlike the time generator, the two spatial operators do not commute in the
non-relativistic setting, hence endowing the spacetime associated with (2+1)-Galilean gravity with a non-trivial quantum structure.
This contrasts the relativistic
version given in~\cite{BHMplb} of the noncommutative spacetime here presented, where the time operator is no longer a central element, a fact which strenghtens the interest in furthering its investigation.  This (2+1)-Lorenztian noncommutative spacetime has `cyclic' commutation relations involving spacetime coordinates (see~\eqref{qa}), that can be considered as a cosmological constant deformation of the pioneering Snyder noncommutative spacetime \cite{Snyder}, which is of (noncommutative but flat) Minkowskian nature.

On the other hand, the non-relativistic contraction scheme developed in the paper rises a couple of remarks. First, a look at Table~\ref{table2} shows that twisted
$\kappa$-deformation models yield trivial non-relativistic spacetimes, as the corresponding $r$-matrices only exhibit contributions containing the two
central extensions. Second, it would be interesting to identify whether these remaining
contracted classical $r$-matrices displayed in Table~\ref{table2} can be associated with other DD structures for the extended (2+1) Galilean algebras.

Finally, a subtle issue in quantum group approaches to quantum gravity is to analyse the possible dynamics unfolded in the corresponding quantum
spacetimes. In this respect, the geometrical interpretation of CS gravity provides some clues as to how massive, spinning particles may be
accounted for in this framework (see, for instance,~\cite{Bernd2,MajSchroers} and references therein). Therefore, a challenging problem is to identify and discuss the  dynamics emerging from the  PL  structures associated with the quantum double of (2+1)-Galilean gravity considered in this paper and its
Lorenztian counterpart given in~\cite{BHMplb}.  This issue should be faced through the same combinatorial quantisation approach followed in~\cite{Bernd1,Bernd2}. The construction of the full quantum  $\overline{\rm NH_{\omega,\xi}(2+1)}$ group here presented can be considered as a first step in this direction, that should be followed by the explicit obtention of its corresponding dual quantum $\overline{\rm NH_{\omega,\xi}(2+1)}$ algebra (the quantum double), along with its unitary representations, star structure and fusion rules, together with its associated quantum $R$-matrix. A glimpse at the results here provided suggests that the derivation of all these structures for the quantum double with non-vanishing cosmological constant will be quite demanding. Nonetheless, the use in this paper from the very beginning of the kinematical basis (that coincides with the one used in the classification of (2+1)-Lorentzian DDs  performed in \cite{BHMcqg}) will hopefully provide a direct interpretation of the results so obtained, as it has been the case for the appearance of the noncommutative mass coordinate in the Galilean noncommutative spacetime~\eqref{spacetime1}. Work on all these lines is in progress.

\section*{Acknowledgements}

This work was partially supported by the Spanish MICINN   under grant   MTM2010-18556.  P.~N. acknowledges a postdoctoral fellowship from Junta de Castilla y Le\'on (Spain).




\end{document}